%% file: ms.tex
\documentclass[preprint]{aastex}
\usepackage{graphics}

\slugcomment{Astrophysical Journal, in press}

\shorttitle{Ly$\alpha$-Emitting Galaxies at z = 2.1}
\shortauthors{Guaita et al.}

\begin{document}

\title{Ly$\alpha$-Emitting Galaxies at z = 2.1 in ECDF-S: Building Blocks of Typical Present-day Galaxies?\footnote{Based on observations obtained at Cerro Tololo Inter-American Observatory, a division of the National Optical Astronomy Observatory, which is operated by the Association of Universities for Research in Astronomy, Inc., under cooperative agreement with the National Science Foundation.}}

\author{Lucia Guaita\altaffilmark{2}, Eric Gawiser\altaffilmark{3}, Nelson Padilla\altaffilmark{2}, Harold Francke\altaffilmark{2}, Nicholas A. Bond\altaffilmark{3}, \\Caryl Gronwall\altaffilmark{4}, Robin Ciardullo\altaffilmark{4}, John J. Feldmeier\altaffilmark{5}, Shawn Sinawa\altaffilmark{4}, \\Guillermo A. Blanc\altaffilmark{6}, Shanil Virani\altaffilmark{7}}

\email{lguaita@astro.puc.cl}    

\altaffiltext{2}{Departmento de Astronomia y Astrofisica, Universidad Catolica de Chile,
    Santiago, Chile}
\altaffiltext{3}{Department of Physics and Astronomy, Rutgers, The State University of New Jersey, Piscataway, NJ 08854}
\altaffiltext{4}{Department of Astronomy\&Astrophysics Penn State University, State College, PA 16802 }
\altaffiltext{5}{Department of Physics and Astronomy, Youngstown State University, Ohio 44555-2001}
\altaffiltext{6}{Department of Astronomy, University of Texas at Austin, Austin, TX 78712}
\altaffiltext{7}{Department of Astronomy, Yale University, New Haven, CT 06520-8101}

\begin{abstract}

We discovered a sample of 250 Ly$\alpha$ emitting (LAE) galaxies at $z\simeq2.1$ in an ultra-deep 3727{\AA} narrow-band MUSYC image of the Extended Chandra Deep Field-South.  LAEs were selected to have rest-frame equivalent widths (EW) $>20$~{\AA} and 
emission line fluxes $F_{Ly\alpha} >2.0 \times 10^{-17}$~erg~ cm$^{-2}$~s$^{-1}$, after carefully subtracting the continuum contributions from narrow-band photometry. 
The median emission line flux of our sample is $F_{Ly\alpha} = 4.2 \times10^{-17}$~erg~cm$^{-2}$~s$^{-1}$, corresponding to a median Ly$\alpha$ luminosity L$_{Ly\alpha} = 1.3 \times10^{42}$~erg~s$^{-1}$ at $z\simeq2.1$. At this flux our sample is $\geq90$~\% complete.
Approximately 4\% of the original NB-selected candidates were detected in X-rays by Chandra, and 7\% were detected in the rest-frame far-UV by GALEX; these objects were eliminated to minimize contamination by AGN and low-redshift galaxies.
At L$_{Ly\alpha} \geq 1.3 \times10^{42}$~erg~s$^{-1}$, the equivalent width distribution is unbiased and is represented by
an exponential with scale-length 83~$\pm10$~{\AA}.  Above this same luminosity threshold, we find a number density of $1.5\pm0.5\times 10^{-3}$~Mpc$^{-3}$.
Neither the number density of LAEs nor the scale-length of their EW distribution show significant evolution from $z\simeq3$ to $z\simeq2$.
We used the rest-frame UV luminosity to estimate a median star formation rate of 4 M$_{\odot}$~yr$^{-1}$. The median rest-frame UV slope, parametrized by the color $B-R$, is that typical of dust-free, 0.5-1 Gyr old or moderately dusty, 300-500 Myr old population.
Approximately 30\% of our sample is consistent with being very young (age~$<100$ Myr) galaxies without dust.
Approximately 40\% of the sample occupies the $z\sim2$ star-forming galaxy locus in the $UVR$ two color diagram, but the true percentage could be significantly higher taking into account photometric errors. 
Clustering analysis reveals that LAEs at $z\simeq2.1$ have $r_0=4.8\pm0.9$ Mpc, corresponding to a bias factor $b=1.8\pm0.3$.  
This implies that $z\simeq2.1$ LAEs reside in dark matter halos with median masses log(M/M$_{\odot})= 11.5^{+0.4}_{-0.5}$, which are among of the lowest-mass halos yet probed at this redshift.   
We used the Sheth \& Tormen conditional mass function to study the descendants of these LAEs and found that their typical present-day descendants are local galaxies with L$^*$ properties, like  
the Milky Way.

\end{abstract}

\section{Introduction}

The search for high-redshift star-forming galaxies advanced rapidly with the introduction of the Lyman Break Galaxies (LBG) technique (\citealt{Gu:1990},  \citealt{StHm:1992}, \citealt{Steidel:1999}) that takes advantage of the lack of flux at wavelengths shorter than the Lyman break at 912~{\AA} ~rest frame due to absorption of ionizing photons by neutral hydrogen, located in stellar atmospheres, in the interstellar medium (ISM), and in the intergalactic medium (IGM) between galaxies.
At $z=3$ the break is located in the observed U band and at higher redshift it moves into optical and infrared bands.
This has allowed an exploration of star-forming galaxies at redshifts $3\leq z\leq8$ via imaging from ground and space (e.g. \citealt{Steidel:2003}, \citealt{Bouwens:2006}, \citealt{Ouchi:2008}) and spectroscopy on 8--10 meter telescopes (e.g. \citealt{Shapley:2001},  \citealt{Shapley:2003}). 
A significant fraction of high redshift LBGs show the Ly$\alpha$ line in emission (\citealt{Shapley:2001}). This emission offers additional information about the process of star formation inside these galaxies and radiative transfer in their ISM. 

Looking for galaxies with Ly$\alpha$ in emission has become an important photometric technique that permits us to find faint (R $\sim$ 27) star-forming galaxies at high redshift. This technique consists of comparing the flux density measured in a narrow-band filter, revealing observed-frame Ly$\alpha$ emission to that found in the broad-band filters, representing the continuum.  
Thanks to the intensity of this emission line, the resulting Ly$\alpha$ emitting (LAE) galaxies provide a special population of high redshift galaxies.
The properties of LAEs have been extensively studied at $z\ge3$ (e.g. \citealt{Ouchi:2005}, Venemans et al. 2005, \citealt{Gawiser:2006b}, \citealt{Gronwall:2007}, ~\citealt{Nilsson:2007}).  LAE samples are composed primarily of galaxies fainter in the continuum than LBGs; Ly$\alpha$ Emitting galaxies therefore probe the lowest bolometric luminosities at high redshift.

Theoretical models, that include radiative transfer inside star-forming galaxies (\citealt{V:2006}, \citealt{SV:2008}, \citealt{V:2008},  \citealt{Atek:2009}), were also developed to understand how Ly$\alpha$ photons form in HII regions and then escape the galaxy, depending on resonant scattering by neutral hydrogen, dust absorption and velocity dispersion in the interstellar medium. 
The amount of dust and the interstellar medium geometry can affect the escape of Ly$\alpha$ photons and hence the shape of the line. 
Clumpy media could permit Ly$\alpha$ photons to escape, even if the galaxy is not dust-free (\citealt{Neufeld:1991}, \citealt{Fin:2008}, \citealt{Fyougold:2009}). 

Spectral Energy Distribution (SED) fitting of the stacked multi-wavelength photometry of $z\simeq3$ LAEs (\citealt{Gawiser:2007}, \citealt{Lai:2008}) shows they are a young (median starburst age of $\sim20$ Myr), low stellar mass (M$\sim10^9$ M$_{\odot}$), modest SFR (median SFR $\sim2$ M$_{\odot}$~yr$^{-1}$), low dust (A$_V\leq$0.2) population of galaxies in an active phase of star formation. SEDs have also shown older population best fits for subsamples of LAEs at $z>3$ (\citealt{Pirzkal:2007}, \citealt{Ono2009}, \citealt{Nilsson:2009}).
SED fitting of individual galaxies showed older ages, higher stellar mass and more dust for continuum-bright LAEs drawn from LBG samples (\citealt{Shapley:2001}, \citealt{Tapken:2007}, \citealt{Pentericci:2009}). \citet{Stiavelli2001} had also shown redder colors for some LAEs at $z\simeq2.4$.
Recently \citealt{Nilsson:2009} presented the first results of observations of LAEs at $z\simeq2.3$, inferring evolution in the properties from $z\sim3$ to $z\sim2$, with more diversity in photometric properties at $z\simeq2.3$.

Clustering analysis of LAEs showed $z\ge4$ LAE bias factors (\citealt{kovac:2007}, \citealt{Ouchi:2004}) expected for progenitors of massive elliptical galaxies in the local Universe, while $z\simeq3.1$ LAEs could be progenitors of L$^*$ galaxies (\citealt{Gawiser:2007}). Semi-analytical simulations were also able to reproduce these results (\citealt{Orsi2008}).  Lower redshift observations, including clustering, will reveal evolution from high to low redshift. For this reason we were motivated to study LAE samples at redshift around 2. This will trace the star formation properties of this type of galaxy at the epoch of the peak of cosmic star formation density (\citealt{Madau:1998}, \citealt{Giav2004}). It also promises to reveal $z\sim0$ descendants of LAEs at $z\simeq2.1$.

In this paper we describe the results from ultra-deep 3727~{\AA} narrow-band MUSYC (MUlti-walengthSurvey Yale Chile, \citealt{Gawiser:2006a}) imaging of the 998 arcmin$^2$ Extended Chandra Deep Field-South. In sections 2 and 3 we present the observations and the data reduction. In Section 4 we summarize the selection of the LAE sample and estimate the possible contaminants. We present the properties of the LAE sample in Section 5: number density, star formation rate, colors and clustering. In Section 6 we discuss the results and derive conclusions.  

We assume a $\Lambda$CDM cosmology consistent with WMAP 5-year results (\citealt{Dunkley:2009}, their table 2), adopting the mean parameters  $\Omega_m=0.26$, $\Omega_{\Lambda}=0.74$, H$_0=70$ km~sec$^{-1}$~Mpc$^{-1}$, $\sigma_8=0.8$.

\section{Observations}
\label{sec:obs}

\begin{figure}[htbp]
\begin{center}
\includegraphics[width=90mm]{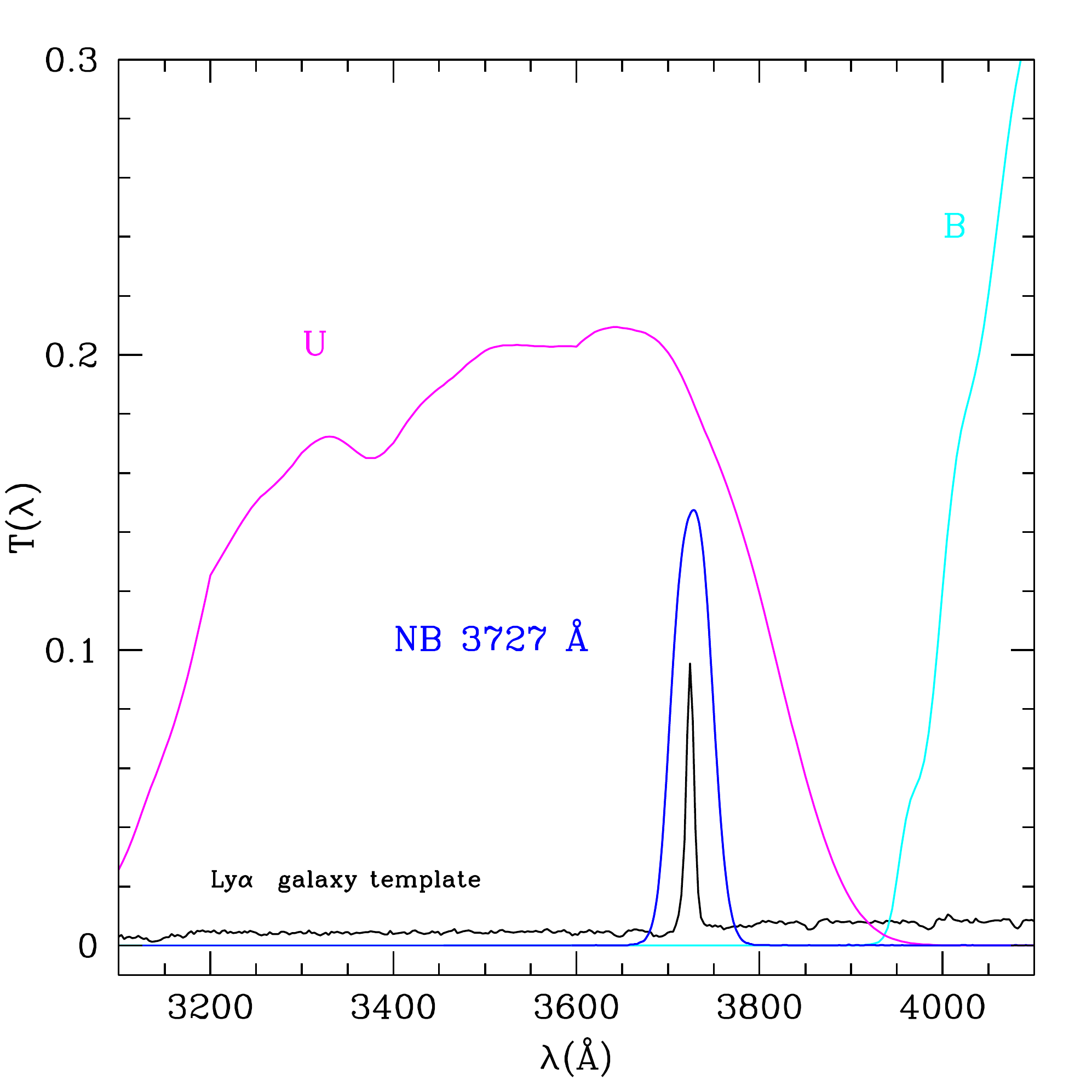}
\caption{Transmission curve of the NB3727 narrow-band filter (blue) at MOSAIC II and of the U (magenta) and B (cyan) broad-band filters at WFI. The filter transmission curves have been multiplied by the detector quantum efficiency and the atmospheric transmission at one airmass. A typical Ly$\alpha$ galaxy template is also shown in black.} 
\label{fig:o2UBfiltershape}
\end{center}
\end{figure}

Our observations of the Extended Chandra Deep Field-South (ECDF-S) were carried out at the CTIO Blanco 4m telescope, using the MOSAIC II CCD camera (eight 2048 $\times$ 4096 CCDs, each with two amplifiers). 
We took advantage of public broad-band $UBVRI$ images of ECDF-S taken with WFI at the ESO 2.2m telescope, processed by the Garching-Bonn Deep Survey (GaBODS, \citealt{Hildebrandt:2006}), and reprojected to match the MUSYC $BVR$ 
image (Table \ref{tab:allbands}).\footnote{These images will be available as part of the MUSYC public data release, labelled v2.}   
We use NB3727 to detect the Ly$\alpha$ emission line flux and a weighted combination of $U$ and $B$ to measure the continuum flux density.  
Fig.~\ref{fig:o2UBfiltershape} shows the transmission curves of these filters.

We used the narrow-band filter with 
response centered at 3727~{\AA} (FWHM=50~{\AA}), originally designed to detect the [OII] emission line, corresponding to the Ly$\alpha$ emission line wavelength at $z=2.07\pm0.02$.    
Our field was imaged during 2007, December $3-13$, using hour-long exposures to avoid the read-out noise limit (see Table \ref{tab:obs}). The total exposure time was about 36 hours and the median seeing of the run was 1.4$"$.  
 The raw images cover a field of view of 36$\times$36 arcmin$^2$, the ECDF-S (central coordinates Right Ascension = 3$^h$32$^m$29$^s$, angular Declination =  -27$^o$48$'$47$"$). 

\section{Data Reduction}
\label{sec:reduction}

The NB3727 narrow-band data were reduced using the IRAF $mscred$ package designed to process MOSAIC frames.  We followed the NDWFS (NOAO Deep Wide-Field Survey) cookbook \footnote {http://www.noao.edu/noao/noaodeep/ReductionOpt/frames.html} 
as modified by \citeauthor{Gawiser:2006a} (2006a) plus a few additional steps described below.  
The principal steps in the reduction process were:\\
i) Creating and applying an improved Bad Pixel Mask (BPM).  To better represent the distribution of bad pixels and columns than the default BPM, we combined together all the twilight flats and all the object frames of the run in the NB3727 filter. We used the $sflatcombine$ task, which takes into account the difference in signal levels (exposure times), to make a median combination of the input frames. The features in the combined frame represent bad pixels and columns that are present in all the frame files.
Applying the $ccdproc$ task to all the raw bias, sky flat and object frames, we replaced the updated version of BPM regions through linear interpolation between good pixels; \\
ii) Improving the World Coordinate System (WCS) information provided in the header of the raw object frames, using interactively the $msccmatch$ task. To estimate the astrometric correction, we built a list of point (stellarity parameter $>0.8$) sources from the MUSYC ECDF-S catalog \footnote {http://www.astro.yale.edu/MUSYC/}, detected in a deep composition of $B$, $V$, $R$ bands. The sources are
uniformly distributed in the field, not saturated, but bright enough ($11<B<21$, $10<R<18$) to be seen in the narrow-band image;\\
iii) Removing cosmic rays (CRs), particularly important given the single frame exposure time of one hour. We used LACOSMIC software package (\citealt{vandokkum2001} \footnote{http://www.astro.yale.edu/dokkum/lacosmic}) with 4 iterations.
Cosmic ray pixels (image features with sharp edges) were replaced by the median of the surrounding $good$ pixels. We chose a contrast limit between CR and underlying object equal to 5, as required for a conservative discrimination between bright stars and cosmic rays and a CR detection limit designed for HST space images. 
These CR pixels are added to the BPM;\\
iv) Transforming the MOSAIC frames into tangent plane projected images with $mscimage$. To be able to stack all the images of the run into one deep image, we used the MUSYC $BVR$ image as a reference to define the tangent point, orientation and the pixel scale of 0.267$''$~pixel$^{-1}$;\\
v) Matching signal levels using $mscimatch$. 
We defined a scaling between each exposure frame, comparing the intensities of a sample of point sources 
from the MUSYC $BVR$ catalog.  
We separated all the hour-long NB3727  
images into two groups of $\sim$18 hours each that we call the first (1H) and second (2H) half of the run. We later used the two halves of the run to search for spurious sources revealed by significant flux variations between
the two halves;\\
vi) Stacking of all the first and second half images, following the point-source-optimized weighting procedure developed by \citeauthor{Gawiser:2006a} (2006a).  
In Table \ref{tab:images} we show the properties of all the images of the run.
As the final step of the image reduction, we applied $mscimatch$ to the two halves to scale them in intensity and then performed a weighted stack of the two halves to create the final NB3727 image of the full run. The overall seeing of the final image is 1.4$''$;\\
vii) Estimating and subtracting the sky background using the Source Extractor program (SExtractor, \citealt{bertin1996}). The sky background was estimated as the average of the background counts in boxes of 64 $\times$ 64 pixels and then the average was median-filtered smoothed across six 64 $\times$ 64 pixel boxes;\\
viii) Shifting and trimming our final stacked image to have the same size and areal coverage as our reference MUSYC $BVR$ and hence the other MUSYC broad-band images, covering 31.6' $\times$ 31.6' at 0.267 arcsec/pixel scale (\citealt{Gawiser:2006a});\\
ix) Normalized to effective exposure time of one second and added
photometric calibration for the final NB3727 image and both halves.
As the photometric calibration was determined using Galactic stars, for extragalactic studies we subtracted off the factor A$_{\lambda} = 0.05$ mags, as appropriate for the near-UV wavelength range and E(B-V) = 0.01 at this location, to account for 
extinction by dust in the Milky Way (\citealt{SFD98}).
Photometric calibration of NB3727 via Landolt standard and spectrophotometric standard stars proved challenging, so we 
adjusted the nominal photometric calibration by 0.35 magnitude to set the median 
$UB$$-$NB3727 color (defined in \S 4) to zero in AUTO photometry. This causes star colors to match those
predicted by \citet{picklesyy} templates to within 0.1 magnitudes.

\section{SAMPLE SELECTION}
\label{sec:sample}
 
We extracted sources following the method described in \citeauthor{Gawiser:2006a} (2006a). 
We used SExtractor to detect and extract sources from the 
final NB3727 image.  
We filtered by the approximate PSF (a 9$\times$9 pixel Gaussian grid with FWHM 5 pixels) and required a minimum of one pixel above the chosen threshold of 0.8 sigma.
We optimized this detection 
threshold to detect the highest number of sources while avoiding a large percentage of spurious ones. We estimated the number of spurious objects, assuming symmetrical background fluctuations, by running SE on the ``negative" of the narrow-band image (narrow-band image multiplied by ``-1") and counting the number of negative detections as a function of our parameters.

 We ran SExtractor in dual mode with the NB3727 narrow band as the detection image and each of the MUSYC broad bands ($UBVRIzJHK$ plus $U38$) and the $1H$ and $2H$ stacks as the measurement images. For each of 19455 sources in the NB3727 catalog, this measured their corresponding fluxes in the other bands. 
We used the corrected-aperture method from \citealt{Gawiser:2006a} to convert optimized-aperture to total APCORR fluxes.  
Most objects in our NB3727 detected catalog have relatively low signal-to-noise.  By comparing with the higher S/N broad-band images, we found an 0.1$"$ rms offset between the narrow-band detection image centroid and the better-determined broad-band centroid, implying a consequent 13\% underestimate of broad-band flux of our catalog objects.  Bright sources do not exhibit these centroiding errors, so it is not a problem of astrometry.  To compensate this loss, we increased the broad-band APCORR fluxes by this amount.
The signal-to-noise for point sources 
in the NB3727 stacked image was optimized using an aperture diameter of 
1.4$''$ which contains 40\% of the signal for point sources. In the case of broad-band images, the optimal aperture had 1.2$"$ diameter, which contains 41\% of $U$ and 43\% of $B$ point source flux.  

\begin{figure}[htbp]
\includegraphics[width=90mm]{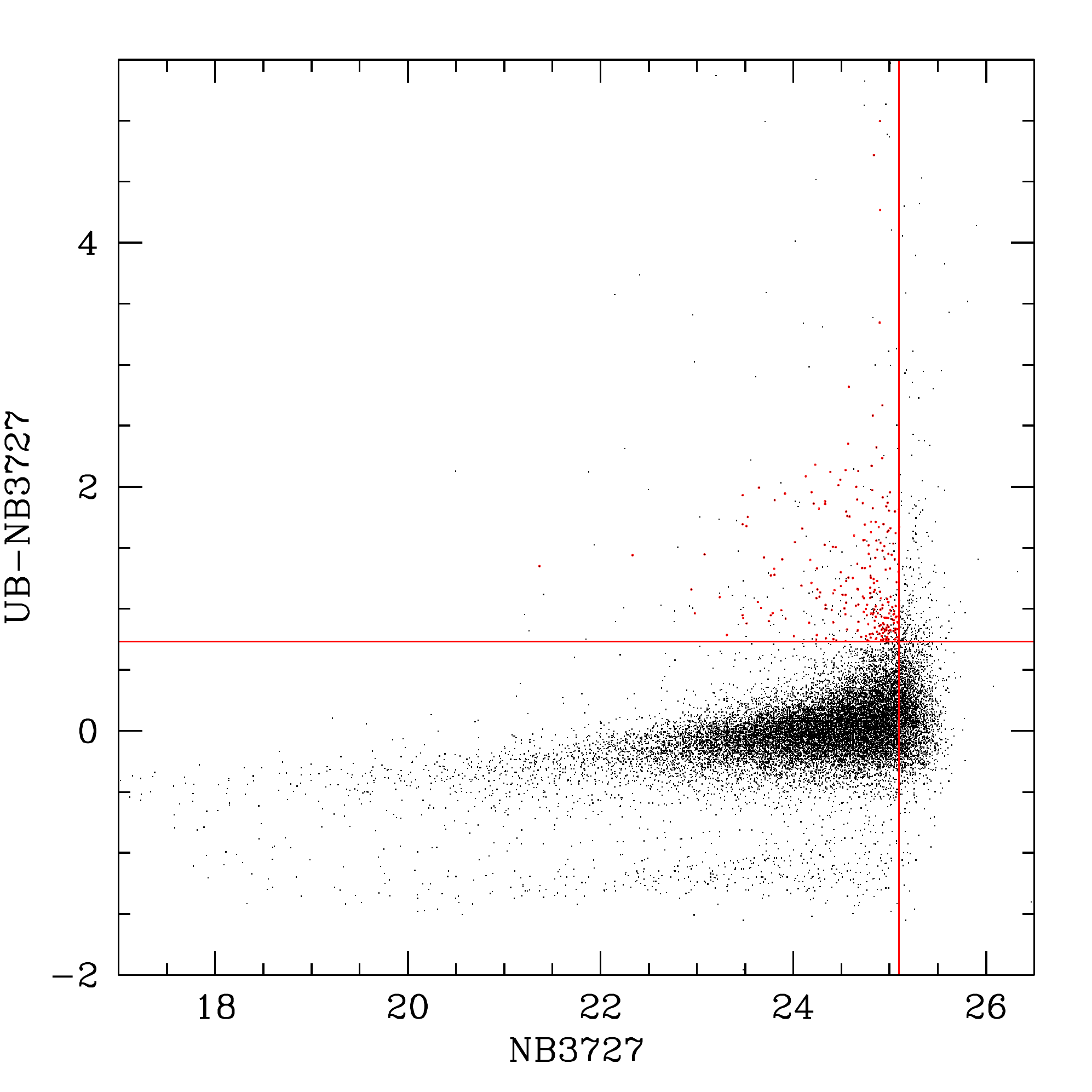}
\includegraphics[width=90mm]{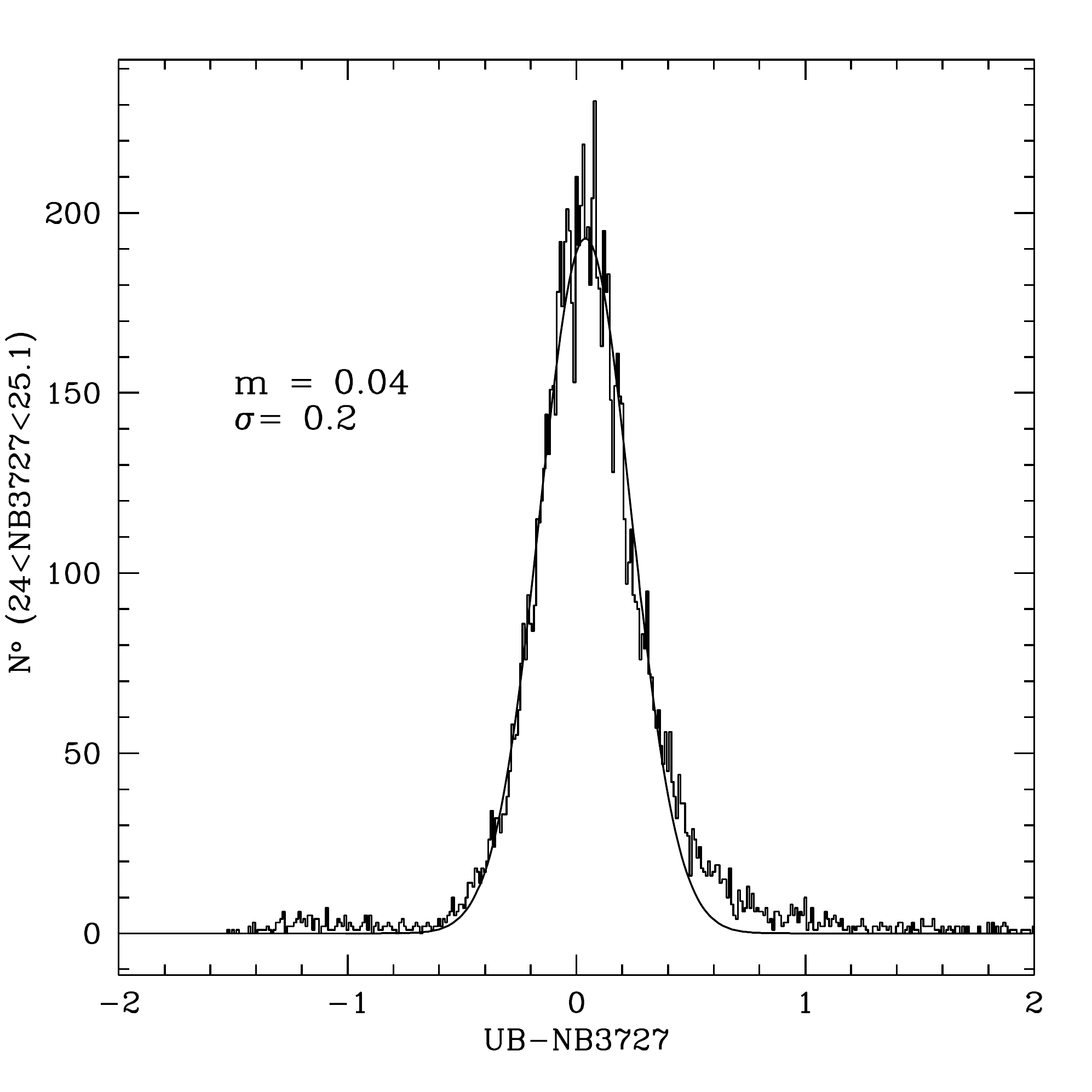}
\caption{{\it Left panel}: UB-NB3727 color versus NB3727 magnitude. Black points represent the total sample of 19455 sources, while the red circles represent the LAE candidates. The horizontal red line represents the cut at rest-frame equivalent width bigger than 20 {\AA} ($UB$-NB3727$>0.73$). The vertical line represent the 5$\sigma$ detection limit of the survey at NB3727~$=25.1$.
{\it Right panel}: UB-NB3727 color distribution for those objects in the 
initial catalog with 24$<$NB3727$<$25.1. 
The Gaussian fit is obtained considering the range -0.5$\leq$UB-NB3727$\leq$0.5. 
The $\sigma$ of the best fit Gaussian curve is 0.2. 
This means that our color cut UB-NB3727$>$0.73 is 
selecting objects above 3.5$\sigma$. 
} 
\label{fig:UBo2}
\end{figure}

\begin{figure}[htbp]
\includegraphics[width=90mm]{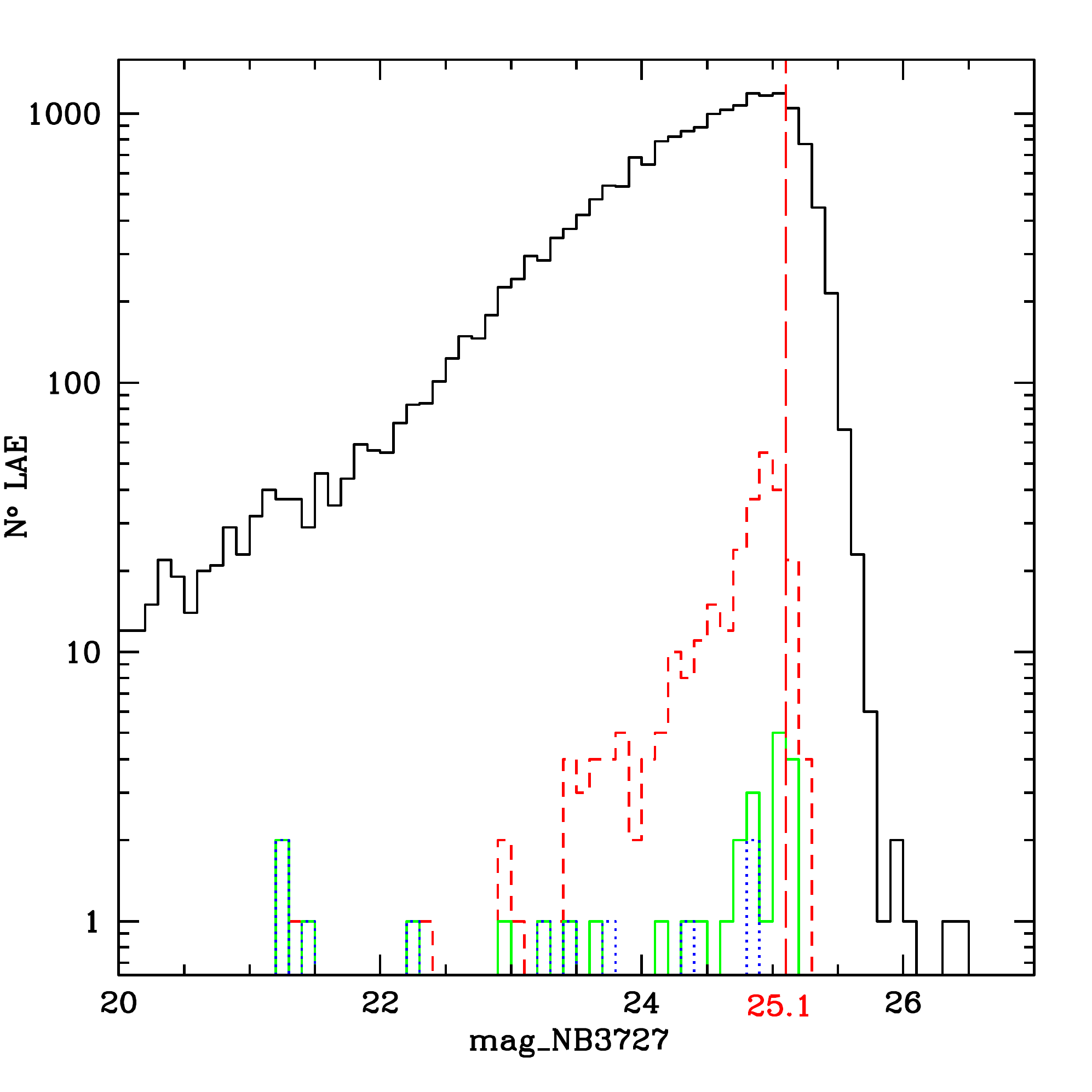}
\includegraphics[width=90mm]{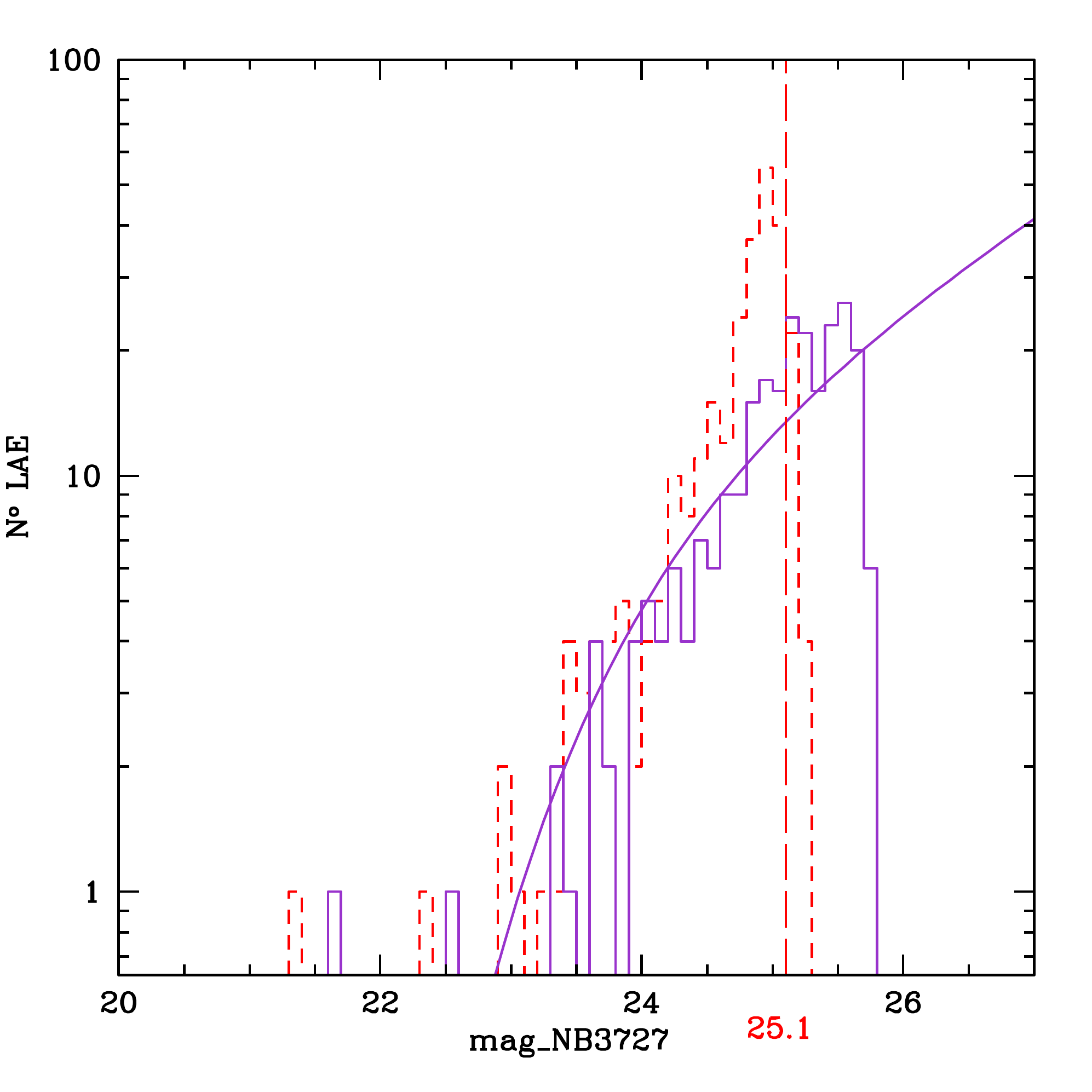}
\caption{{\it Left panel}: NB3727 magnitude distribution of catalog 19,455 objects found in the initial catalog, plotted as the black solid histogram, compared to the objects 
 satisfying the LAE selection criteria 
(excluding the NB3727~$<25.1$ criteria), plotted as the red
dashed line. 
Objects that would have satisfied the LAE selection criteria but have X-ray 
(GALEX)  detections are shown in blue dotted (green solid) line.
The vertical long-dashed red line represents the 5$\sigma$ detection limit 
of NB3727=25.1.
{\it Right panel}: The NB3727 magnitude distribution of the LAEs with NB3727 
magnitude brighter than 25.1 in red dashed line. 
The distribution of the magnitudes of the same sample after continuum 
subtraction is shown in violet solid line. The violet curve represents the magnitude distribution derived from 
the $z\simeq3.1$ LAE luminosity function, moved to $z\simeq2.1$ 
assuming no luminosity evolution. }  
\label{fig:mag}
\end{figure}

LAE candidates were selected with the following criteria: \\
1.  {\it Narrow-band detection at 5$\sigma$ significance}.  We chose objects with magnitudes brighter than the typical 5$\sigma$ NB3727 detection limit of magnitude 25.1, corresponding to Ly$\alpha$ line fluxes, F$_{Ly\alpha} > 2.0 \times 10^{-17}$~erg~sec$^{-1}$~cm$^{-2}$ and luminosities, L$_{Ly\alpha}> 6.4 \times 10^{41}$~erg~sec$^{-1}$, after carefully subtracting the continuum contributions from narrow-band photometry (see Appendix for details, equations (\ref{a})-(\ref{i})). 16,872 objects of our catalog satisfy this ``global" signal-to-noise criterion.\\ 
2. {\it Local signal-to-noise $>5$.}  Our analysis of detections in the negative image indicated that the global S/N criterion would still leave 33 fake sources.  These fake sources are concentrated in the region of the amplifier with the highest readout noise.  Even though SExtractor uses a semi-local measurement of the background rms as a detection threshold, we found that applying an additional cut of the ratio between the aperture flux and the photometric error on it bigger than 5, $f_\mathrm{aper} / \sigma_{f_\mathrm{aper}} > 5$, eliminated all but 7 of the detections in the negative image.  Implementing this criterion, most of the excluded objects are located in the region of that noisiest amplifier, that would otherwise have been classified as LAEs. 15,882 objects satisfy the first and this second criteria.\\
3.  {\it Narrow-band excess corresponding to EW$>20${\AA} }.  We defined a color $UB-$NB3727 as the difference in magnitudes between the
$UB$ and NB3727 flux densities (see Appendix, equations (\ref{u}),(\ref{v}),(\ref{colorequation})), 
where $UB$ refers to the linear combination of $U$ and $B$ flux densities, $f_{UB}=0.8f_U + 0.2f_B$, motivated by the central wavelengths of the filters. 
A positive value of $UB-$NB3727 indicates an excess 
in the narrow-band  flux density.  In order to obtain an Equivalent Width (EW) rest-frame cut of EW $>$20 {\AA} (\citealt{Gronwall:2007}, \S 5.2 of this paper), we required UB-NB3727$>$0.73. 
(Fig.~\ref{fig:UBo2}). This generated an initial list of 367 LAE candidates. \\
4. {\it 1$\sigma$ significance of the narrow-band excess versus a pure continuum spectrum.}  We required 
\begin{equation}
f_{NB3727} - f_{UB} > \sqrt{ \sigma^2(f_{NB3727})+\sigma^2(f_{UB})}
\label{1s}
\end {equation}
 to avoid contamination by continuum-only objects whose narrow-band photometry fluctuated upwards or continuum photometry fluctuated downwards due to Poisson statistics.  While this only requires the presence of a narrow-band flux density excess at $1\sigma$ significance, combined with the requirement of 
$UB-$NB3727$>0.73$, it appears to avoid most such contaminants, at the cost of some incompleteness as discussed further in \S \ref{sec:contamination}.  It also has the benefit of eliminating a number of objects with poor photometry from the sample by virtue of their larger photometric uncertainties.  
Most objects that passed the previous criteria, but were eliminated by this one, are faint (AB magnitude NB3727$\sim$24.0) and extended (NB3727 half light radius $>$1.4$''$).  
They consist of multiple objects in the deepest $BVR$ image that are blended  by the larger NB3727 PSF into single faint objects centered  between the $BVR$ object positions. In this case, aperture photometry at the NB3727 centroid underestimates the continuum flux, leading to a false narrow-band excesses.  Because the APCORR pipeline includes an extended object correction and flux uncertainty increase based upon the half light radius, these objects have large enough uncertainty in their narrow-band flux excess to be eliminated by this criterion. 48 objects are excluded after including this requirement, leaving 319 objects.\\
5.  {\it Lack of variation in narrow-band flux between first and second half stacks}.  We exclude four objects for which
\begin{equation}
|f_{1H}-f_{2H}| > 3 \sqrt{\sigma_{1H}^2 + \sigma_{2H}^2} 
\label {variability}
\end{equation}
where $f_{1H}$ and $f_{2H}$ correspond to an object's flux density in the stacked NB3727 images of the first and second halves of the run and $\sigma$ represents the uncertainty on each.\\
As the two halves of the run are separated by only a few nights, even AGN are unlikely to show measurable variability on these time-scales. 
 Hence this is primarily a method for eliminating objects whose narrow-band excess appears spurious, 
perhaps coming from a single image due to an 
incompletely subtracted cosmic ray or from a contiguous set of images due to a systematic flaw in bias subtraction or flat-fielding. After this correction, 315 objects remain.\\ 
6.  {\it No saturated pixels.}  We exclude objects satisfying the above criteria that had a maximum SExtractor flag$\geq 4$, implying either uncorrected bad pixels in o2 (4 objects), detections too close to the image border to trust (2 objects), or continuum magnitude bright enough to saturate in at least one band (0 objects at this stage) leaving 309 objects.\\
7.  {\it Not consistent with cross-talk contamination from a bright star.}  The electronic coupling of adjacent amplifiers on MOSAIC II is a serious obstacle for narrow-band excess searches, as a number of the couplings produce echoes that have the same dithering pattern as real objects. A careful analysis of bright star positions versus locations of narrow-band excess sources determined the cross-talk offset to be $\pm2100\pm10$ pixels in declination and $0\pm10$ pixels in right ascension. These offsets were used to generate a cross-talk mask that excluded 15 of our original LAE candidates with only 2 such matches expected by chance.  Visual inspection and analysis of the EW of these objects implies that the vast majority were indeed spurious, so this masking should cause negligible incompleteness in our sample. After the exclusion of these 15 cross-talks, 294 objects remain in the list.\\  
8.   {\it Not detected by Chandra.}
 In addition to Ly$\alpha$ emission at $z\simeq2.1$, a  strong narrow-band excess at $3727${\AA} can be generated by AGN activity.  AGN can show strong emission lines in 
Ly$\alpha$, N V 1240, Si IV 1400, C IV 1550, He II 1640, [C III] 1909, Mg II 2800, and Mg I 2852, all of which could trigger a narrow-band excess. 
Our filter is narrow enough to miss some of the contribution of emission lines broader than $\sim4000$~km~s$^{-1}$, but this still leaves both broad and narrow emission lines as a likely source of AGN contamination.  
Given the deep Chandra imaging available in this field (2 Ms exposure in CDF-S and 250 ks exposure in ECDF-S), we expect to detect X-rays from all unobscured and some obscured AGN at $z\leq 2.1$. 
Therefore we exclude 10 (4\%) NB-selected candidates that we find also in the combined Chandra catalog (\citealt{Luo:2008},  \citealt{Virani:2006}, \citealt{Lehmer:2005}) within a $2''$ radius. This number is significantly bigger that the 1 match expected by chance, meaning that the matching program found real X-ray sources.  These sources are characterized by 21$<R<$25. Excluding the candidates with X-ray detection, 284 objects remain.\\
9.  {\it Not detected by GALEX in NUV or FUV.}  Objects at $z\simeq2.1$ should be invisible in these GALEX filters due to the Lyman break at $\lambda<2800${\AA}, which precisely matches the red cutoff of the $NUV$ filter.  To minimize contamination from low redshift objects, we therefore exclude 24 candidates with detection in one or both GALEX bands within a search radius of 3 arcsec; 4 of those belonged already to the Chandra catalog. Up to 7\% of the  LAEs candidates seem to present a counterpart in the GALEX catalog, quantity consistent with the 30 matches expected by chance, but, in any case, we decided to treat those objects as a separate sub-sample. Their magnitude distribution follows the shape of that of all the selected LAE candidates (Fig.~\ref{fig:mag}). Excluding also the candidates with a counterpart in the GALEX catalog, 264 objects remain in the list.\\
10.  {\it Passed visual inspection.}  The final step in determining our sample of $z\simeq2.1$ LAEs was to visually inspect the NB3727, $U$, $B$, and $BVR$ images of each candidate to ensure that none displayed obvious systematic flaws in object detection or photometry that were missed by the above criteria. Only 14 objects were eliminated at this stage, due to flaws in their photometry caused by source blending that would create biased estimations of narrow-band excess. Most of these were cases of 2 BVR-detected objects blending into one in NB3727, as described above. Keeping the candidate selection process automated except for this final step enables Monte Carlo simulations and will hopefully make our estimates of contamination and incompleteness more secure than if we made widespread, subjective use of visual inspection.  Note that in the analysis of \citet{Nilsson:2009} a visual inspection phase generated a ``maybe'' set of $\sim100$ objects that were excluded from analysis despite not presenting obvious flaws; our approach is the opposite, which has significant advantages for achieving completeness.  We will discuss possible sources of contamination in the next section. Our final sample consists of 250 $z\simeq2.1$ LAEs in 998 arcmin$^2$.

\subsection{Contamination Estimates}
\label{sec:contamination}

In our sample analysis we considered four possible remaining sources of contamination:\\
1.  {\it Spurious objects}, manifesting as pure NB3727 emitters with zero continuum, which causes significant fractional uncertainties on the broad-band flux densities. 
We conducted detailed simulations of false object detection by using SExtractor to search for objects in the ``negative" image defined above using identical detection 
parameters.  This predicts that our sample of LAEs includes 7 false objects, all of them with 24$<$NB3727$<$25.1, $UB$-NB3727 $>0.7$, $UB$-NB3727 $>0$ at 1$\sigma$. 
An empirical analysis was also performed. Since false objects detected in NB3727 have zero continuum, photometric errors should push half to positive and half to negative flux densities in our deepest continuum image, $BVR$, due to the symmetry of fluctuations.  
Hence finding 1 LAE with negative flux density in $BVR$ yields a best estimate of 2 spurious objects. 
Combining these two approaches and the following discussion in the Appendix, we estimate contamination by 4$^{+3}_{-2}$ spurious sources.
We found counterparts in the GEMS  \footnote {http://www.mpia-hd.mpg.de/GEMS/gems.htm} HST-ACS $V$-band images
for 90\% of our LAEs.  Since LAEs are selected via emission-line excess, no continuum is required, but this analysis
does set an upper limit of 10\% for our contamination by spurious objects (which should not have GEMS counterparts).
Similarly we found a 70\% counterpart match rate between $z\simeq2.1$ LAEs and the MUSYC $BVR$ catalog, which are not as deep as GEMS $V$-band and therefore place a weaker constraint on contamination by spurious objects.\\
2. {\it Continuum only objects} that show an NB3727 excess due to photometric noise. 
We assume that continuum-only contaminants are in the range $24<$~NB3727~$<25.1$, as the few brighter candidates would have good photometry.
We fit with a Gaussian curve the distribution of the $UB$$-$NB3727 color for the original catalog of objects with $24<$~NB3727~$<25.1$ (Fig. \ref{fig:UBo2}b). As the Gaussian $\sigma$ is equal to 0.2, we are selecting objects above 3.5$\sigma$ using our color cut. Comparing the ratio between the integrated area at $UB$$-$NB3727~$>0.73$ and under the Gaussian curve in the range $-0.5 \leq UB$-NB3727~$\leq 0.5$,  and accounting for uncertainties in the Gaussian fit, we estimate that $5^{+10}_{-3}$ contaminants belong to the sample selected via $UB$$-$NB3727~$>0.73$.  \\
3.  {\it Lower redshift emission line galaxies} i.e., [O II] emitters.
We expect virtually none of these objects to contaminate our sample, due to their tiny number density at rest-frame EW$>20${\AA} (\citealt{Hogg:1998}) and the small volume available for z$\simeq$0 objects. Local Universe [OII] emitters would be several arcsec across, so would stand 
out clearly in our catalogs.  In any case the exclusion of GALEX detected sources should rule out this contribution.\\   
4.  {\it Obscured AGN}, which are capable of triggering a narrow-band 
excess through their narrow emission lines.  Since we found 10 AGN 
as X-ray sources in the Chandra catalog and some of those may be obscured or Compton thick, and most models predict a roughly equal number of obscured and unobscured AGN at this redshift (\citealt{Treister2004}), we set an 
upper limit on residual AGN contamination of 10$\pm$10 objects.  This will be probed 
via follow-up spectroscopy.  Note that heavily obscured AGN may not show any emission lines at all and therefore would not be found in our sample; this reinforces confidence in our upper limit. 
We stacked 66 LAEs in our sample with coverage in the 2Ms CDFS image (Luo et al. 2009) and found 3$\sigma$ upper limits for the soft-band (hard-band) stacked flux of 4$\times10^{-18}$ (2 $\times 10^{-17}$) erg s$^{-1}$ cm$^{-2}$, corresponding to a luminosity of 1.3 $\times 10^{41}$ (6.7$\times 10^{41}$) erg s$^{-1}$ at $z=2.1$. The observed soft-band implies
a 3$\sigma$ upper limit on the average SFR of 30 M$_{\odot}$ yr$^{-1}$ (Ranalli et al. 2003).   Compared to our typical rest-UV SFR of 4 M$_\odot$ yr$^{-1}$ this implies that the dust correction must be less than a factor of seven.  Because individual X-ray detections above the 2Ms flux limit of 2$\times 10^{-17}$ erg s$^{-1}$ cm$^{-2}$ were removed from our LAE sample in this region, any AGN remaining must have soft-band luminosity below 7$\times 10^{41}$ erg s$^{-1}$.  In the extreme case, 20\% of our sample could contain low-luminosity AGN just below this threshold; this provides a weaker constraint on AGN contamination than those mentioned above. 

Combining all of these sources of contamination 
we expect 19$^{+23}_{-15}$ interlopers 
in our final sample of 250 objects.  Taking the uncertainties into account we estimate the contamination fraction to be $7\pm7$\%.

\section{RESULTS}
\label{sec:results}

In our observation of ECDF-S Lyman Alpha Emitters at $z\simeq2.1$, we achieve the same 5$\sigma$ detection limit in Ly$\alpha$ luminosity (log(L(Ly$\alpha$))=41.8) as the sample of LAEs at $z\simeq3.1$ (Gronwall et al. 2007). They found 154 LAEs in a total area of 992 arcmin$^2$, imaging the ECDF-S with the narrow-band filter at 4990 {\AA} of the MOSAIC II instrument at the 4m CTIO telescope. 
This corresponds to a number density of $1.5\pm0.3 \times 10^{-3}$ Mpc$^{-3}$. They reached a narrow-band magnitude depth NB4990$=$25.4, that corresponds to a Ly$\alpha$ flux limit of $1.5\times10^{-17}$~erg~cm$^{-2}$~s$^{-1}$.  Gawiser et al. (2007) used the same sample to derive spectral and clustering properties of the $z\simeq3.1$ LAE population. 
Fig.~\ref{fig:mag} shows the narrow-band magnitude distribution of our catalog of 19455 objects and the sample of 250 LAEs.
Using the estimate of the continuum at 3727 {\AA} flux (Appendix, equation (\ref{h})), we constructed the distribution of the NB3727 magnitude after subtracting the contribution of the continuum emission, 
also shown in the figure. 
This latter quantity represents the Ly$\alpha$ emission-line flux.
The 5$\sigma$ detection magnitude limit of 25.1 corresponds to an emission line flux 
$F_{Ly\alpha} = 2.0\times10^{-17}$~erg~cm$^{-2}$~s$^{-1}$, 
assuming that the LAE has EW = 20 {\AA} and is at $z=2.066$, where the Ly$\alpha$ emission line 
receives the maximum NB3727 throughput.  Since most emission lines have higher EW and receive lower narrow-band throughput, this is a strong lower limit on the 
Ly$\alpha$ fluxes, and we expect significant incompleteness near this flux. The median flux of our sample is $F_{Ly\alpha} = 4.2\times10^{-17}$~erg~cm$^{-2}$~s$^{-1}$, and the corresponding median Ly$\alpha$ luminosity is $L_{Ly\alpha} = 1.3\times10^{42}$~erg~s$^{-1}$ at $z=2.066$.

\subsection{Number density of LAEs and AGN}
\label{sec:number}
 
We estimate both the catalog and our sample to be 
$\sim$50\% complete at the limiting magnitude of NB3727=25.1 and to be 90\% complete at NB3727$=$24.8. We determined these photometric limits by adding artificial stars to our survey fields in groups of 2000, and repeating until the limits were well defined (1,680,000 artificial stars in all). We therefore estimate 30$\pm$10\% incompleteness for the sample as a whole. The candidates excluded for having GALEX counterparts appear no different in their magnitude distribution (Fig.~\ref{fig:mag}) and match the expected number of chance coincidences with the large GALEX catalog. We therefore expect that excluding these 24 objects has caused $\sim$10\% additional incompleteness for a total of 40$\pm$10\%.   
Because our filter shape matches that used by \citeauthor{Gronwall:2007} (2007), we follow their analysis.
Assuming the same effective filter width as in \citeauthor{Gronwall:2007} (2007) (80\% of the FWHM), we estimate a comoving volume of 124500 Mpc$^3$ in $\Delta~z=$0.033 ($z=$2.082-2.049).  Therefore the number density of $z\simeq2.1$ LAEs to our selection limits is estimated to be 250/124500 = ~$2.0 \times 10^{-3}$ Mpc$^{-3}$.
Given the 7\% contamination estimated above, this suggests a factor of 0.93/0.6 = 1.5 correction to our nominal number density for the sample as a whole. 
We derive a total number density at NB3727~$<25.1$ (corrected for incompleteness) of $3.1\pm0.9\times 10^{-3}$ Mpc$^{-3}$ at $z\simeq2.1$, for which the errors are calculated as the sum in quadrature of the uncertainties in the incompleteness factors, the sample variance due to large scale structure for this volume (\citealt{Somerville:2004}) equal to $\sim$25\%, and the Poisson error. In the total survey area our number density corresponds to a surface density of  
$0.4\pm0.1$ LAEs arcmin$^{-2}$. The number density of LAEs at $z\simeq2.1$ can also be defined as $12\pm4$ arcmin$^{-2}$ per unit of redshift. \citealt{Gronwall:2007} calculated $4.6\pm0.4$ arcmin$^{-2}$ per unit of redshift.
 
For comparison with models and other surveys, it is critical to measure the number density of LAEs above a fixed Ly$\alpha$
luminosity limit.  At the lowest Ly$\alpha$ luminosities in our survey, there is a strong selection effect, with only low-EW objects able to
make the NB3727~$<25.1$ cut due to their continuum contribution to the narrow-band photometry.  However, above the Ly$\alpha$ luminosity limit of $1.3\times10^{42}$~erg~s$^{-1}$ (Fig. \ref{fig:ew}a), the sample has no selection effect on EW and is $>$~90\% complete and we calculate a number density (corrected for incompleteness) of $1.5\pm0.5\times 10^{-3}$ Mpc$^{-3}$. Restricting the $z\simeq3.1$ LAE sample \citeauthor{Gronwall:2007} (2007) to this same luminosity limit, its number density becomes $1.1\pm0.2\times 10^{-3}$ Mpc$^{-3}$. This corresponds to an evolution factor of $1.4\pm0.5$ from $z\simeq3.1$ to $z\simeq2.1$. We reach twice as deep a Ly$\alpha$ luminosity limit as \citeauthor{Nilsson:2009} (2009).  They selected their sample at $z\simeq2.3$ at a 5$\sigma$ detection limit of 25.3 magnitudes in a 3$"$ aperture diameter, using a FWHM$=$129 {\AA} filter. Restricting our sample to match their luminosity limit of 2.8 $\times 10^{42}$ erg~ sec$^{-1}$, we find a number density of $0.65 \pm 0.2 \times 10^{-3}$ Mpc$^{-3}$, consistent with their $0.62 \times 10^{-3}$ Mpc$^{-3}$, which was also corrected for incompleteness.

In the volume of our survey we found 10 X-ray sources (4\% of our NB-selected catalog) within a search radius of 2$"$, optimized to avoid random matches; 4 of them were also found in the GALEX catalog.  If all of these objects lie at $z\simeq2.1$, this implies a number density of Ly$\alpha$-detected AGN of $8.0\times 10^{-5}$ Mpc$^{-3}$, but since an unknown fraction of these objects are at other redshifts this is an upper limit.  At $z\simeq2.3$ Nilsson et al. (2009) found 13\% (private communication) of their candidates to be X-ray sources detected by Chandra using a search radius of 5$"$.   This initially sounds like a disagreement with our "AGN fraction" of 4\%, which does not change when we use a 5$"$ Chandra search radius.  However, we note that the number of X-ray sources selected via narrow-band excess by Nilsson et al. (2009) corresponds to a consistent number density of $\sim$10$^{-4}$ Mpc$^{-3}$ under the same unlikely assumption that all of the objects lie at $z\simeq 2.1$.   Restricting our sample to the 2x brighter luminosity limit of Nilsson et al. (2009), we find that the percentage of X-ray sources increases to 10\%, so the results are fully consistent.  Because X-ray detected narrow-band excess objects are found selectively on the bright end of the narrow-band magnitude distribution, the inferred number density is far more useful than the percentage given the variations in Ly$\alpha$ luminosity limit between surveys.

\subsection{Equivalent Width distribution} 
\label{sec:ew}

As described in the Appendix, equation (\ref{colorequation}), the $UB$$-$NB3727 color is related to the observed-frame equivalent width (EW) of the Ly$\alpha$ line of the galaxy, via a relation that depends on the total filter transmission curves.
We used this to solve for EW given observed $UB$$-$NB3727 colors. 
  As we can see from the left panel of the Fig. \ref{fig:ew}a for log(L(Ly$\alpha))\geq42.1$ the sample is unbiased in the sense of equivalent width versus Ly$\alpha$ luminosity. In fact the 5$\sigma$ detection limit selection, represented by the solid lines in the figure, requires that faint objects in Ly$\alpha$ luminosity (log(L(Ly$\alpha))<42.1$) have low equivalent widths (EW mostly less than 50 {\AA}), so that the sum of their continuum and emission-line contributions gives them sufficient narrow-band flux density. 
For this reason we restrict the sample to the brighter half to build the EW distribution. 
The distribution of the rest-frame EW ($=\frac{EW_{obs}}{(1+z)}$) of the brighter candidates is represented in Fig.~\ref{fig:ew}b as a black histogram.
We fit the distribution with an exponential law ${\rm d}$N/${\rm d}$EW=N exp$^{-EW/W_0}$, that represents the best fit. In the same figure we also show an exponential law as used in Gronwall et al. (2007) (dashed cyan curve) and  Nilsson et al. (2009) (orange dotted curve).  
Fixing the normalization to produce the right total number of objects, we get a best-fit exponential scale of w$_0=83^{+10}_{-10}$ {\AA}.  This characteristic equivalent width is comparable to that measured at $z\simeq3.1$ by Grownwall et al. (2007), w$_0=76^{+11}_{-8}$ {\AA}, but it is greater than the value measured by Nilsson et al. (2009) at $z\simeq2.3$, w$_0=48.5\pm1.7$ {\AA}.
For a continuum-selected population of galaxies, for example LBGs, we expect objects with Ly$\alpha$ either in emission, in absorption or with no line, in a roughly Gaussian distribution of EW centered at zero (Shapley et al. 2003). For this reason, we also fit the distribution of equivalent width with a Gaussian function ${\rm d}$N/${\rm d}$EW=N exp$^{-EW^2/2 \sigma^2}$, truncated at EW$>20${\AA} and found a best fit 
Gaussian centered at zero with $\sigma_{gauss}=90^{+10}_{-10}$ (reduced $\chi^2 =$ 1.05, calculated with Poisson errors).
However, the exponential is a better fit (reduced $\chi^2$ = 0.9). We compare this Gaussian fit with that calculated by  Ouchi et al. (2008) at $z\simeq3.1$. Our $\sigma_{gauss}$ value is smaller than their vale of $\sigma_{gauss}=130\pm10$, implying in average smaller EWs for the objects in our brighter half of the sample. This result is also related to a possible evolution from $z\simeq5.7$ ($\sigma_{gauss}=270$) as they claim. 

\subsection{Star Formation Rates} 
\label{sec:sfr}

As indicated by Kennicutt (1998), in the range 1500-2800 {\AA} the UV continuum is nearly flat in L$_{\nu}$ and is a good estimator of the star formation rate:
\begin{equation}
SFR(UV) = 1.4 \cdot 10^{-28} \cdot L_\nu (1500-2800 {\AA}) (erg/sec/Hz).
\end{equation}
This assumes a constant SFR over timescales longer than the lifetime of the dominant UV emitting population, at least 10$^8$ years
a Salpeter IMF and that $L_\nu$ has been corrected for dust extinction. 
Spectral Energy Distribution (SED) fitting of typical LAE spectra at $z\simeq3.1$ \citep{Gawiser:2007} shows that dust is negligible in most LAEs, which are observed in a nearly $dust-free$ phase of star formation. 
We assume here that no dust correction is necessary, making our UV SFRs formally lower limits.  
We used the $R$ band flux density at $\sim$ 2000 {\AA} as the estimator of the $z\simeq2.1$ LAE rest-frame UV continuum via
\begin{equation}
L_\nu (UV) (erg~sec^{-1}~Hz^{-1}) = f_{\nu, R} (\mu \mathrm{Jy}) \cdot 10^{-29} \cdot \frac{4\pi D_L^2}{(1+z)} ,
\label{eq:l_nu}
\end{equation}
where D$_L$ is the luminosity distance at $z\simeq2.1$.
Using different rest-frame UV flux estimators, such as $B$ or $V$ band,  
we observed differences in SFR values of up to 20 \%.
 
From recombination line estimators and scaling H$\alpha$ relation, it is possible to calculate the SFR from Ly$\alpha$ emission line luminosity:
\begin{equation}
SFR(Ly\alpha)=9.1 \cdot 10^{-43} \cdot L(Ly\alpha)(erg/sec),
\end{equation}
where $L(Ly\alpha)$ is the integrated luminosity in the Ly$\alpha$ emission line in ergs s$^{-1}$,
\begin{equation}
L(Ly\alpha) = f_{\nu, NB} \cdot10^{-29}\cdot4\pi D_L^2\cdot \frac{\int (c/\lambda^2) T_{NB}(\lambda) d\lambda}{T_{EL}}  .
\label{eq:L_lya}  
\end{equation}
Here, $T_{EL} = T(\lambda_{EL})$ is the transmission of the NB filter at the wavelength of the emission line, where the expected value is $<T_{EL}>$ (Appendix, equation (\ref{b})) and $f_{\nu, NB}$ is the flux in $\mu$Jy in the NB3727 narrow-band filter after subtracting the continuum (see \S2).  

Fig.~\ref{fig:sfr} compares the SFRs measured from UV and Ly$\alpha$. The reduced density of objects at the upper left (SFR(UV)$>10$ M$_{\odot}$ yr$^{-1}$) and lower left (SFR(UV)$<2$ M$_{\odot}$ yr$^{-1}$ and SFR(Ly$\alpha$)$<1$ M$_{\odot}$ yr$^{-1}$) of the plot is at least partially caused by our rest-frame EW$>$20{\AA} and 5$\sigma$ detection limit selections.
Due to resonant scattering of neutral hydrogen, Ly$\alpha$ photons are preferentially absorbed by dust.  Hence the ratio between the SFR estimated from UV continuum and Ly$\alpha$ emission can give an indication on the dust content of typical LAEs at $z\simeq2.1$. The median of the ratio for the objects of the sample with fluxes above the 90\% completeness is $\sim1.5$, consistent with the value found for LAEs at $z\simeq3.1$ (\citealt{Gronwall:2007}). However, we observe a scatter around these median values, due to photometric errors, mainly at faint $R$ band magnitudes, or intrinsic galaxy diversity. A forthcoming spectral energy distribution analysis (Guaita et al. 2010, in preparation) will reveal typical galaxy properties, such as dust, age and SFR more precisely. 
So far our best estimation of the typical SFR of the sample is from the UV estimator, median SFR(UV) equal to $4.0\pm0.5$ M$_{\odot}$ yr$^{-1}$. This is a moderate value of SFR, in agreement with the SED results derived at $z\simeq3.1$ (Gawiser et al. 2007). 

\subsection{Rest-Ultraviolet Colors}
\label{twocolor}

Fig.~\ref{fig:BR} shows $R$ as a function of the $B-R$ color and the distribution of $B-R$ colors of our sample of 250 objects. 
In this Figure, we also plot the median $B-R$ color with error bars showing the median uncertainty in this color for bins of width 0.5 mags in $R$-band.   
The scatter is bigger than the photometric errors for $R<25$, but is comparable for $25<R<27$. The part of the plot with $R>27$ is occupied by few objects consistent with being
pure emission line objects.

The majority of LAEs are blue. We see an almost constant scatter in $B-R$ as a function of $R$. Also, as the photometric errors are smaller at brighter magnitude and comparable to the observed scatter at the faint end, there is a larger intrinsic scatter in $B-R$ at brighter $R$.
The distribution of objects in the $R$ vs $B-R$ plot shows a relatively uniform occupation of the $-0.5<B-R<1$ range. The median $B-R$ color of the sample is 0.16, for a subsample of $R<25$ LAEs the median is 0.38, and for the subsample of $R\geq25$ it is 0.07.
There are a few very bright objects ($R>$ 24) that occupy a red tail of the $B-R$ color distribution. These are characterized by log(L(Ly$\alpha))<42$.

\citet{Gronwall:2007} found that the median $R$ band magnitude of the $z \simeq 3.1$ LAE sample is 27, fainter than the 
$R = 25.5$ detection limit of Lyman Break Galaxies (LBG, Steidel et al. 2003). Similarly, the median $R$ magnitude (Fig.~\ref{fig:uvr}a) of our $z \simeq 2.1$ LAE sample is 25.3, meaning that roughly half of our LAEs could be selected as ÒBXÓ star-forming galaxies (SFGs) by the criteria 
of Steidel et al. (2003). However, this overlap further depends upon the rest-UV ($UVR$) continuum colors of the galaxies. By subtracting the contribution of the emission line from the $U$ band magnitude (Appendix, $f_{\nu,U,only~ continuum}$), we generate the pure continuum $Ucorr-V$ color. Fig.~\ref{fig:uvr}b shows the two-color diagram, $Ucorr-V$ vs $V-R$.
The solid lines delimit the LBG region (upper polygon) and 
the ``BX'' region corresponding to SFGs at $2\leq z \leq2.7$ (central polygon). 
These regions were generated using the \citeauthor{Bruzual:2003} (2003) code, 
assuming a constant star formation rate and a range of ages between 1Myr and 2Gyr . 
We simulated colors for the MUSYC filter transmission curves, 
including a dust extinction law (Calzetti 2000) 
parametrized by $0<E(B-V)<0.3$ and absorption by the IGM \citep{Madau:1995}.

The median $U_{corr}-V$ color of the sample is 0.7, while the median $V-R$ color is  0.12. Hence the typical LAE at $z\simeq2.1$ is located in the lower part of the selection region of BX galaxies, as expected given the $2\leq z \leq2.7$ range of the latter. 
Indeed, 40\% of $R<25.5$ LAEs at $z\simeq2.1$ occupy the BX region, with more scatter for galaxies with fainter continuum (Fig.~\ref{fig:uvr}b). This is the challenge of the narrow-band technique; we expect to find emission lines from continuum faint, therefore less massive, SF galaxies. 
84/250 objects in our sample meet the BX colors in $UVR$ and 60/250 meet both the colors and the typical magnitude requirement of $R<25.5$.  

\subsection{Clustering analysis}
\label{sec:clustering}

We calculated the angular correlation function (Fig.~\ref{fig:cluster}b) of our sample of candidates distributed as in Fig.~\ref{fig:cluster}a and, after projecting it, the correlation length, $r_0$, and bias factor following \citeauthor{Francke:2008} (2008). 

The angular correlation function, $\omega(\theta)$, was calculated using the Landy \& Szalay (\citealt{Landy:1993}) estimator. We used a random catalog of one hundred times the number of our observed data objects to minimize Poisson noise in the calculation of random-random pairs. 
The observed angular correlation function was deprojected to the spatial correlation function $ \xi(r) = (\frac{r}{r_0})^{-\gamma}$, following \citet{Simon2007}.
The fit to the angular correlation function was performed in a two-step manner: first, the double integral of the redshift distribution was calculated (in comoving radial distance scale) and tabulated as a function of $\theta$ and $\gamma$. Then the function $\omega$\_ideal($\theta$, r$_0$, $\gamma$) was formed by multiplying by the r$_0^{\gamma}$ factor (fixing $\gamma=1.8$). The fitting function is "$\omega$\_ideal - IC", where IC represents the integral constraint, IC= $\int(\omega(\theta$) RR($\theta$) d$\theta$ = 0.05681.
Finally the fitting function $\omega$\_model = $\omega$\_ideal - IC was fitted to
the estimated correlation function $\omega_{Landy \& Szalay}$, using $\chi^2$.

We corrected for the contamination factor estimated in \S4 as the contribution of unclustered contaminants. As the contamination rate is so low, the presence of clustered contaminants would make little difference. The uncertainty in the contamination estimate (7\%) has been propagated into the error bar for r$_0$ and added in quadrature to its total error budget. We found $r_0= 4.8\pm0.9$ Mpc, fitting $\theta$ from 40 to 600". This was chosen to avoid the 1-halo term at small scales and to avoid sampling noise at big scales. 
In Fig.~\ref{fig:cluster}a we can observe hints of a large-scale inhomogeneity in the spatial distribution of the LAE candidates at $\delta>-27.75$ and RA~$>53.1$.  We are in the process of confirming via spectroscopy the candidates in that region. We find that the correlation lengths calculated including or excluding these candidates are consistent and their only effect on the angular correlation function can be found at scales $\sim$720", outside the angular range of our fit.
In order to compare our result to other galaxy populations, we used the Sheth \& Tormen (1999) conditional mass function to predict the expected bias evolution as a function of redshift (Fig. 9). The bias evolution tracks plotted in this diagram were calculated from the median of the mass distribution of descendants for a family of dark matter halo masses at high redshift.  The dashed lines correspond to conditional mass function trajectories for bias evolution from Sheth \& Tormen theory. These curves are drawn starting at effective bias values of 2,3,4,5,6,7,8 and 9 at a redshift of 6.0, corresponding to halo populations with median masses of log(M/M$_{\odot}$) = 8.4, 9.7, 10.4, 10.9, 11.3, 11.6, 11.9, and 12.1, respectively at that epoch. 
The bias factor represents the amplitude of galaxy over-densities versus those of dark matter and it is our preferred quantity for clustering strength comparisons.
In the same figure we show the measured values of bias factor for LAEs and other star-forming galaxies as a function of redshift.  Green circles represent the bias values calculated for this sample of LAEs and that from Gronwall et al. (2007) at $z\simeq3.1$. LAEs were observed to be the least clustered population at $z\sim$3 \citep{Gawiser:2007} with a bias factor b~$=1.9^{+0.4}_{-0.5}$. 

In this survey, we measured a bias factor b~$=1.8\pm0.3$ for our sample of LAEs at $z\simeq2.1$, corresponding to a median dark matter halo mass of log(M/M$_{\odot})= 11.5^{+0.4}_{-0.5} $ for the population. 
Using the estimation of the mass function from \citeauthor{Shetormen:1999} (1999), the number density of the $z\simeq2.1$ halos of that median mass is 7.2$^{+19.2}_{-4.5} \times 10^{-3}$ Mpc$^{-3}$, about four times smaller than what we calculated at $z\simeq3.1$ ($30^{+250}_{-23} \times10^{-3}$ Mpc$^{-3}$).
So the occupation fraction, calculated by the ratio between the number density of LAEs and the number density of the halo population, rises from the $5^{+10}_{-4}$ found at $z\simeq3.1$ to $43^{+115}_{-30}$ at $z\simeq2.1$, due to the increase in the LAE number density, although the increase is not statistically significant given the large uncertainties.
Following the conditional mass function tracks to $z=0$, the interesting result is that LAEs at $z\simeq2.1$ appear to be progenitors of present-day  L$^*$ galaxies.

\section{DISCUSSION AND CONCLUSIONS}
\label{sec:discussion}

We imaged the ECDF-S using a NB3727 narrow-band filter, corresponding to the wavelength of Ly$\alpha$ emission at $z\simeq2.1$. 
Following the formalism described in the Appendix, we applied the color cut $UB$$-$NB3727$ > 0.73$ 
and additional significance criteria that yielded a sample of 250 LAEs.
In our observation we achieve the same 5$\sigma$ detection limit in Ly$\alpha$ luminosity (log(L(Ly$\alpha$))=41.8) as the sample of LAEs at $z\simeq3.1$ (Gronwall et al. 2007, Gawiser et al. 2007). Therefore we are able to look for indications of evolution between $z\sim$2 - 3. Concentrating on $z\sim2$, we compare LAEs with  star-forming galaxies (Steidel's BX sample), which can also show the Ly$\alpha$ line in emission. In many cases our analysis concentrates on the typical properties of the LAE sample as a whole; it is important to remember that there will always be cases of individual LAEs whose physical properties differ considerably from those of the typical LAE. 

The magnitude distribution of LAEs at $z\simeq2.1$ (Fig.~\ref{fig:mag}) is consistent with that predicted by the $z\simeq3.1$ LAE Ly$\alpha$ luminosity function, but with about twice the normalization, i.e. total number density. As reported in \S 5.1 we calculated a LAE number density of $3.1\pm0.9\times 10^{-3}$ Mpc$^{-3}$, taking into account the estimated incompleteness of the sample, an evolution in the number density of a factor of $2.1\pm0.7$ versus $1.5\pm0.3 \times 10^{-3}$ Mpc$^{-3}$ reported by \citeauthor{Gawiser:2007} (2007) at $z\simeq3.1$. Our number density is consistent with the value, found by Nilsson et al. (2009) at $z\simeq2.3$ when we restricted our analysis to objects matching their $\sim2 \times$ brighter luminosity limit. At the Ly$\alpha$ luminosity limit, at which the sample is complete, we calculate a number density of $1.5\pm0.5\times 10^{-3}$ Mpc$^{-3}$, that implies an increasing factor of $1.4\pm0.5$, consistent with that calculated for all the sample.

We derive the equivalent width distribution (\S5.2), representative of the $z\simeq2.1$ LAE sample in Fig.~\ref{fig:ew}. As we can see in Fig.~\ref{fig:ew}a, for log(L(Ly$\alpha))\geq42.1$ the sample is unbiased in the sense of rest-frame equivalent width versus Ly$\alpha$ luminosity. We consider the unbiased brighter half of the sample to build the histogram in Fig.~\ref{fig:ew}b. Fitting this distribution with an exponential law, this is consistent with that from Gronwall et al. (2007) for the sample at $z\simeq3.1$ and broader than that found at $z\simeq2.3$ by Nilsson et al. (2009). In Fig.~\ref{fig:ew} we associated the value EW~$=400$~ {\AA} to the objects characterized by an unphysical equivalent width (Appendix, equation \ref{colorequation}). The objects with $EW_{rest-frame}>250$ present $UB>27$. 
Most of the objects in the sample with $EW_{rest-frame}<50$ also have log(L(Ly$\alpha))<42.1$, meaning that their continuum flux boosted them above the narrow-band catalog detection limit. This behavior was less prevalent at $z\simeq3.1$ by Gronwall et al. (2007), although the 5$\sigma$ detection limit creates a similar trend, as shown by the blue curve in Fig.~\ref{fig:ew}a.
As it is described in the Appendix, we estimate the observed EW from the observed color $UB$$-$NB3727. Those estimations are in perfect agreement with those obtained from continuum flux density and Ly$\alpha$ emission line flux. As described in Dayal et al. (2009), the measured EW at the border of the galaxies can be increased by the cooling of collisionally interstellar medium excited HI atoms, while the continuum almost remains unchanged, but intergalactic medium absorption can attenuate Ly$\alpha$ flux and so decrease the observed EW. 

The Ly$\alpha$ luminosity reveals star formation activity inside a galaxy (\S5.3).  Log(L(Ly$\alpha))=42.1$, the median Ly$\alpha$ luminosity of our sample, corresponds to SFR(Ly$\alpha$) =  1.2 M$_{\odot}$~yr$^{-1}$, as indicated by the dashed-dotted line of Fig.~\ref{fig:sfr}. In the same figure we observe the range of SFR(UV) values. The median
LAE at $z\simeq2.1$ has a moderate SFR(UV) of $\sim$ 4 M$_{\odot}$~yr$^{-1}$. The ratio of $\sim$1.5 in the values of SFR(UV)/SFR(Ly$\alpha$), for the unbiased half of the sample, is caused by potentially complex radiative transfer of Ly$\alpha$ photons in the dusty, possibly clumpy interstellar medium inside the galaxies \citep{Atek:2009}. Given the overlap in clustering bias it is worth considering whether $z\simeq2.1$ LAEs could populate the low (stellar) mass tail of continuum-selected star-forming galaxies at  $z\sim2$. 
We find that the LAE SFR(UV) is 10 times lower than that calculated from UV continuum and H$\alpha$ line emission by \citeauthor{Steidel:2004} (2004) for star-forming galaxies at $z\sim2$. 
The Kennicutt estimator, used to derive the star formation rate from UV continuum, assumes that the galaxy is at least 10$^7$ yr old with  roughly constant SFR.

We find (\S5.4) that 240/250 (96\%) of $z\simeq2.1$ 
LAEs are blue ($B-R$)$<1$, with 73/250 (30\%) having ($B-R$)$<0$.  This is in good agreement with the $z\simeq3.1$ sample in both criteria.  
In fact at $z\simeq3.1$, LAEs with $R<25$ have median color $B-R=0.53$ (\citealt{Gronwall:2007}).
Our result agrees with the findings of \citet{Nilsson:2009} at $z\simeq2.3$ in the fraction of LAEs having ($B-R$)$>0$, but their conclusion that most LAEs are ``red'' depended on considering all objects with rising spectra in $f_{\nu}$ to be red. 
A reasonable split of galaxies into blue and red is achieved by using ($B-R$)$=1$ as the dividing line, and we suspect that the sample of \citet{Nilsson:2009} will show similar properties when this is applied. In fact looking at their Fig. 4 and deriving the behavior of the color $B-R$ from the slope $\beta(B-R)$, we see that their galaxies are essentially blue, based on our definition. 
 
 The appearance of bimodality in the LAE rest-UV color at $R<25$ is intriguing.
The blue branch is presumably dominated by young, dust-free star-forming galaxies, since unobscured (blue) AGN should have been eliminated from our sample due to their X-ray emission.  The red branch may contain obscured (dust-reddened) AGN, galaxies with Ly$\alpha$ emission from recent starbursts but an overall older or dustier stellar population and low-redshift interlopers that will be identified via follow-up spectroscopy.  
We calculated the evolutionary tracks of galaxies at z$\sim$2 in the $U-B$ vs $B-R$ plane, generated using the GALAXEV (Bruzual \& Charlot 2003) code for a constant star formation rate and a range of masses from 25 Myr to 1 Gyr,  parameters consistent with LAE SED fits.  We see that a 500 Myr old galaxy with dust absorption A$_V=0$ has color $B-R=0$.
If it is star forming, the $U-B$ color, corrected for IGM absorption, is also close to zero. Increasing the age the color $B-R$ becomes slightly bigger than 0. However increasing the amount of dust, for example to A$_V\sim$1, typical for reddened LBG, the star-forming galaxy can assume $B-R$=0.5-0.6. The color $B-R$=1 is achieved by galaxies with significantly more dust than that measured for typical star-forming populations.
There is a smaller difference in $B-R$ between young ($<5 \times10^8$ yr) and old ($>5 \times10^8$ yr) star-forming populations than the difference produced by the increasing reddening.
The observed median($B-R$)~=~0.16 is typical of star-forming galaxies with A$_V$~=~0 and ages of 0.5-1 Gyr or can be consistent with moderate A$_V$ and age 300-500 Myr. Approximately 30 \% of our sample with negative $B-R$ color is consistent with being very young (age$<$100 Myr) galaxies without dust.

We divide in bins of 0.5 magnitude in $R$ and construct Table \ref{tab:Rchar}, which shows the magnitude range, the median color, EW, SFR from Ly$\alpha$ and SFR from the UV continuum. These values are transformed into intrinsic ones, taking into account the dust and gas amount (parametrized by stellar E(B-V) and E$_g$(B-V) ) and radiative transfer effects. 
 The median colors lie inside the ``BX" region except for the faintest bins which have large photometric uncertainties.
 
As expected we observe that the EW values are bigger for the objects that are fainter in the continuum. We calculated $EW_{rest-frame}>250$ for objects with $UB>27$. Statistical fluctuations related to such a faint continua can produce an over-estimation of the equivalent width of these objects. We observe that bright-continuum objects ($UB<24.5$) are also bright in Ly$\alpha$ luminosity.
 For low-EW LAEs ($UB-NB3727$ just $\simeq 0.73$), as expected, the SFR(UV) is significantly bigger than the SFR(Ly$\alpha$). In the table we also report the standard deviations in the $R$ magnitude bins. In the last column the scatter error is less meaningful, because of the proportionality between $R$ flux density and SFR(UV). It is seen that the scatter is as big as the corresponding quantity. In $B-R$ color it is consistent with that was observed in Fig.~\ref{fig:BR}. 

The clustering analysis (\S5.5) gives information about the LAEs at $z\simeq2.1$ as a population and their evolution to redshift zero. 
In Fig.~\ref{fig:cluster2} we see that LAEs at very high redshift ($z>4$, Ou sign, H09 sign) can evolve into massive LBG at $z\sim$3 and also reach, in the local Universe, the bias factor typical of elliptical massive galaxies, corresponding to luminosity between 2.5 and 6.0 L* (as indicated by the points in the figure) and halo masses greater than $4.47 \times 10^{13} $~M$_{\odot}$.  Looking at $z\sim3$ \citep{Gawiser:2007}  LAEs were observed to be blue galaxies and to be characterized by lower clustering than other galaxy samples at that redshift. They can evolve into star-forming galaxies at $z\sim2$ (A0 sign) and then to L$^*$ galaxies in the local Universe.
At  $z\simeq2.1$ we calculate a bias factor b~$=1.8\pm$0.3 for our sample of LAEs. This value is consistent with that found using the conditional mass function for progenitors of  L$^*$ galaxies in the local Universe. It is also consistent with the value calculated for the subset of ``BX" galaxies dimmest in $K$-band ($K_{Vega}>21.5$, \citealt{Adelbergerb:2005}); that is low mass galaxies.
This clustering result matches that of dark matter halos with median masses of log(M/M$_{\odot})= 11.5^{+0.4}_{-0.5} $, which are some of the lowest halo masses  probed at this redshift. 
Our result shows that $z\sim2$ LAEs could also be descendants of $z\simeq3.1$ LAEs, depending on how long dust-free star formation occurs and on possible cyclical repetitions of star formation phases.   
As LAEs at $z\simeq2.1$ are consistent with being the progenitors of present-day and L$^*$ galaxies at $z=0$, they are likely building blocks of local galaxies with properties similar to the Milky Way  and median halo mass  $\geq 2\times 10^{12}$ M$_{\odot}$.

\acknowledgments

We acknowledge helpful conversations with Steven Finkelstein, Peter Kurczynski, Cedric Lacey, Sangeeta Malhotra, Kim Nilsson, Laura Pentericci, Naveen Reddy, James Rhoads, Bram Venemans, Yujin Yang and the unknown referee for the very useful comments on the paper. We thank the anonymous referee for her/his very helpful suggestions that improved the paper.
We are grateful for support from  Fondecyt (\#1071006),
Fondap 15010003, Proyecto Conicyt/Programa de Financiamiento Basal para Centro Cient'ficos y Tecnol—gicos de Excelencia (PFB06),
Proyecto Mecesup 2 PUC0609, ALMA-SOCHIAS fund for travel grants.  
This material is based on work supported by the National Science 
Foundation under grant AST-0807570 and AST-0807885, by the Department of 
Energy under grant DE-FG02-08ER41560 and DE-FG02-08ER41561, 
and by NASA through an award issued by JPL/Caltech. 
E.G. thanks the Berkeley Center for Cosmological Physics and the 
Aspen Center for Physics for hospitality during the 
preparation of this paper.
L.G thanks Rutgers University for hosting her during collaborative research.

{\it Facilities:} \facility{Blanco}

\appendix

\section{Appendix: Calculation of Equivalent Width}
\label{sec:appendix}

\begin{figure}[htbp]
\begin{center}
\includegraphics[width=80mm]{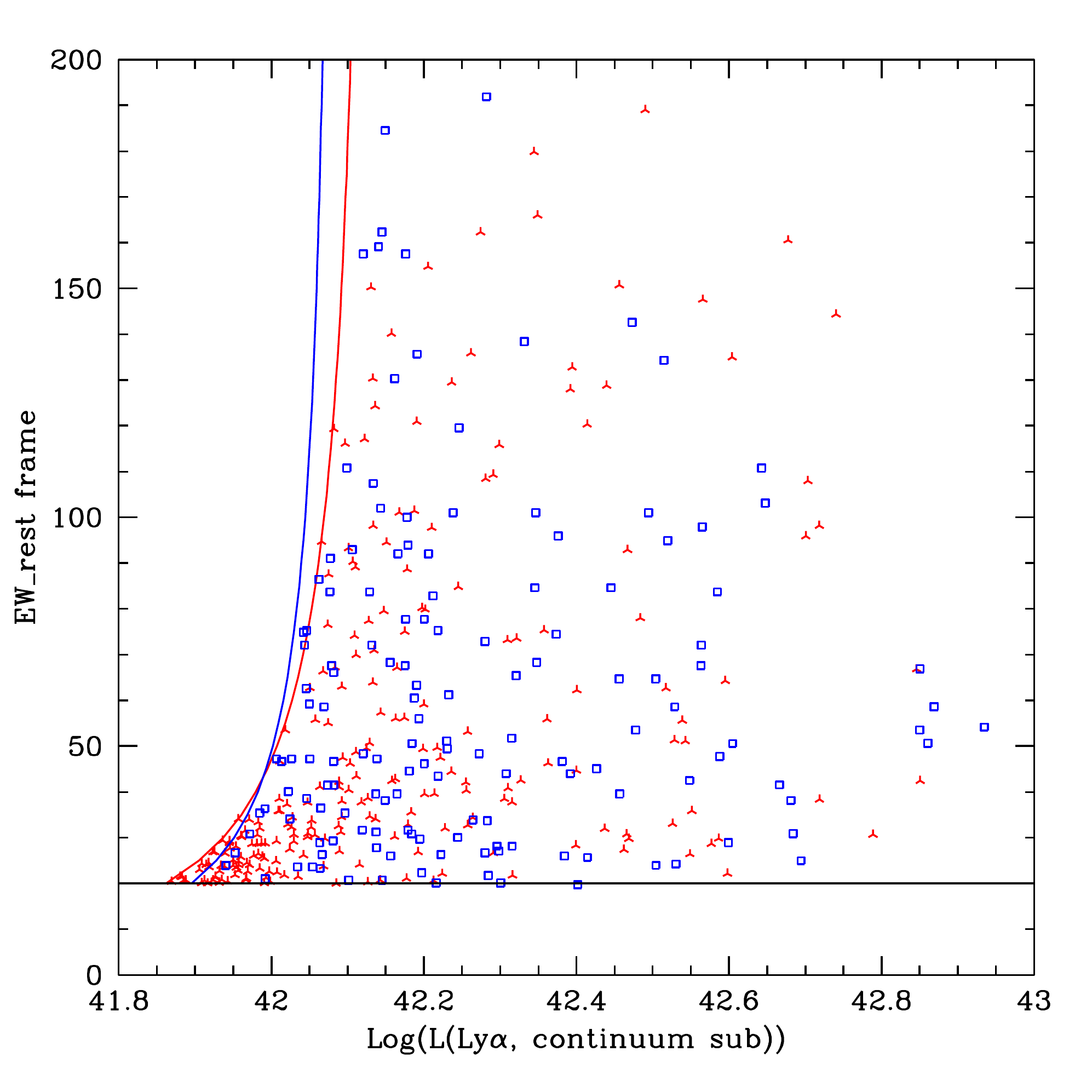}
\includegraphics[width=80mm]{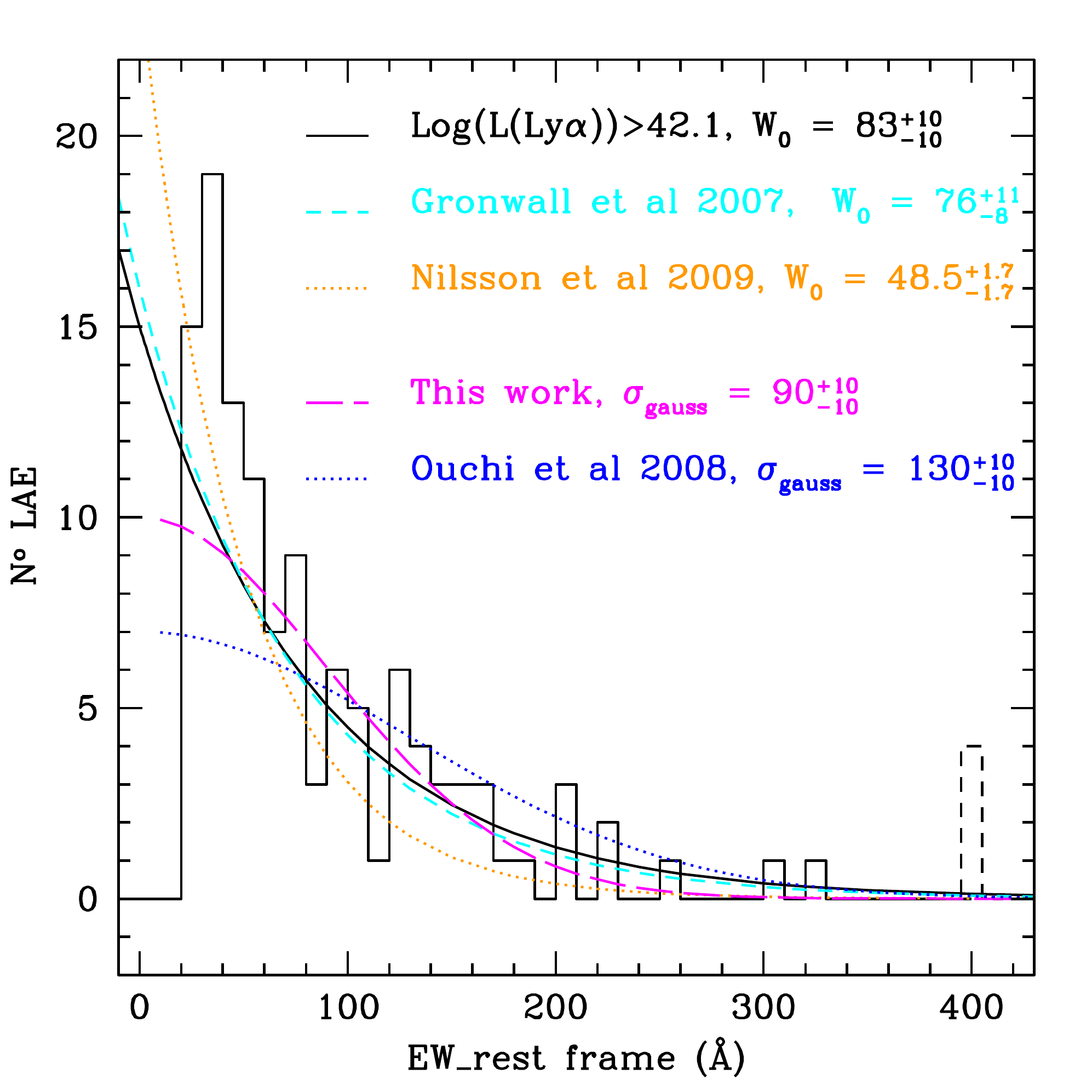}
\caption{{\it Left panel}: 
Rest-frame equivalent width (EW) versus logarithmic Ly$\alpha$ luminosity for our sample of 250 $z\simeq2.1$ LAEs in red triangles and the sample of 154 $z\simeq3.1$ LAEs in blue squares. The solid lines represent the 5$\sigma$ detection limit in magnitude of the samples, in red for the $z\simeq2.1$ LAE survey and in blue for the $z\simeq3.1$ survey. At log(L(Ly$\alpha))\geq 42.1$ the $z\simeq2.1$ sample appears to be unbiased in the sense of rest-frame EW versus Ly$\alpha$ luminosity. The black horizontal line represents the cut at rest-frame EW~$>20$ {\AA}.
{\it Right panel}: Distribution of rest-frame equivalent widths for the $z\simeq2.1$ LAEs with log(L(Ly$\alpha))\geq 42.1$. The black solid line represents its exponential best-fit.
In cyan dashed line we show the best-fit exponential for $z\simeq3.1$ LAEs from Gronwall et al. (2007) and in orange dotted line we show the best-fit exponential for $z\simeq2.3$ LAEs from \citet{Nilsson:2009}.  
The magenta long-dashed line represents the Gaussian fit of the histogram. As a comparison we show the Gaussian fit by Ouchi et al. (2008) as the lower blue dotted line.  
The bin at EW~$=400$ {\AA} includes formally infinite EW estimations i.e., objects whose continuum photometry is dimmer than should be possible even for a pure emission line, 
although this can be caused by photometric noise.} 
\label{fig:ew}
\end{center}
\end{figure}

\begin{figure}[htbp]
\begin{center}
\includegraphics[width=100mm]{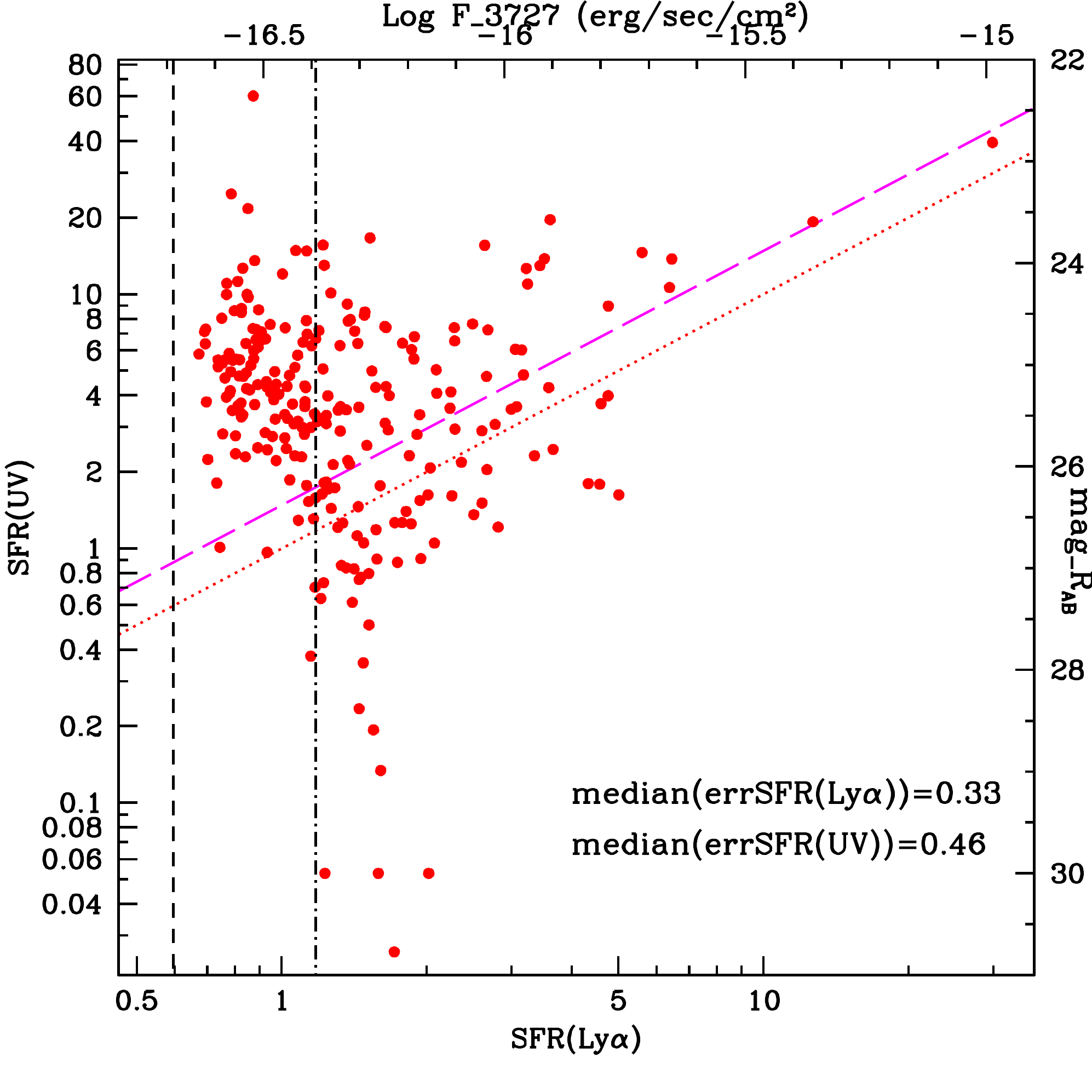}
\caption{SFR(UV) versus SFR(Ly$\alpha$), also labeled with $R$ mag versus monochromatic NB3727 flux for our sample of 250 LAEs.
The vertical dashed line represents the limit flux of the survey (estimated using the maximum transmission filter value and rest-frame EW$=$ 20 {\AA}), while the dash-dotted line represents 90\% completeness limit. The diagonal dotted line represents the case in which SFR(UV)$=$SFR(Ly$\alpha$). The magenta dashed line represents the case in which the SFR(UV) is equal to the SFR(Ly$\alpha) \times$ the median-ratio(SFR(UV)/SFR(Ly$\alpha)=1.5$), for the part of the sample that is complete. The points at $R=30$ represent objects for which the measured flux in $R$ band is negative. The reduced density of objects at the upper left (SFR(UV)$>10$ M$_{\odot}$yr$^{-1}$) and lower left (SFR(UV)$<2$ M$_{\odot}$ yr$^{-1}$ and SFR(Ly$\alpha$)$<1$ M$_{\odot}$ yr$^{-1}$) of the plot is at least partially caused by our rest-frame EW$>$20{\AA} and 5$\sigma$ detection limit selections.}
\label{fig:sfr}
\end{center}
\end{figure}

\begin{figure}[htbp]
\begin{center}
\includegraphics[width=70mm]{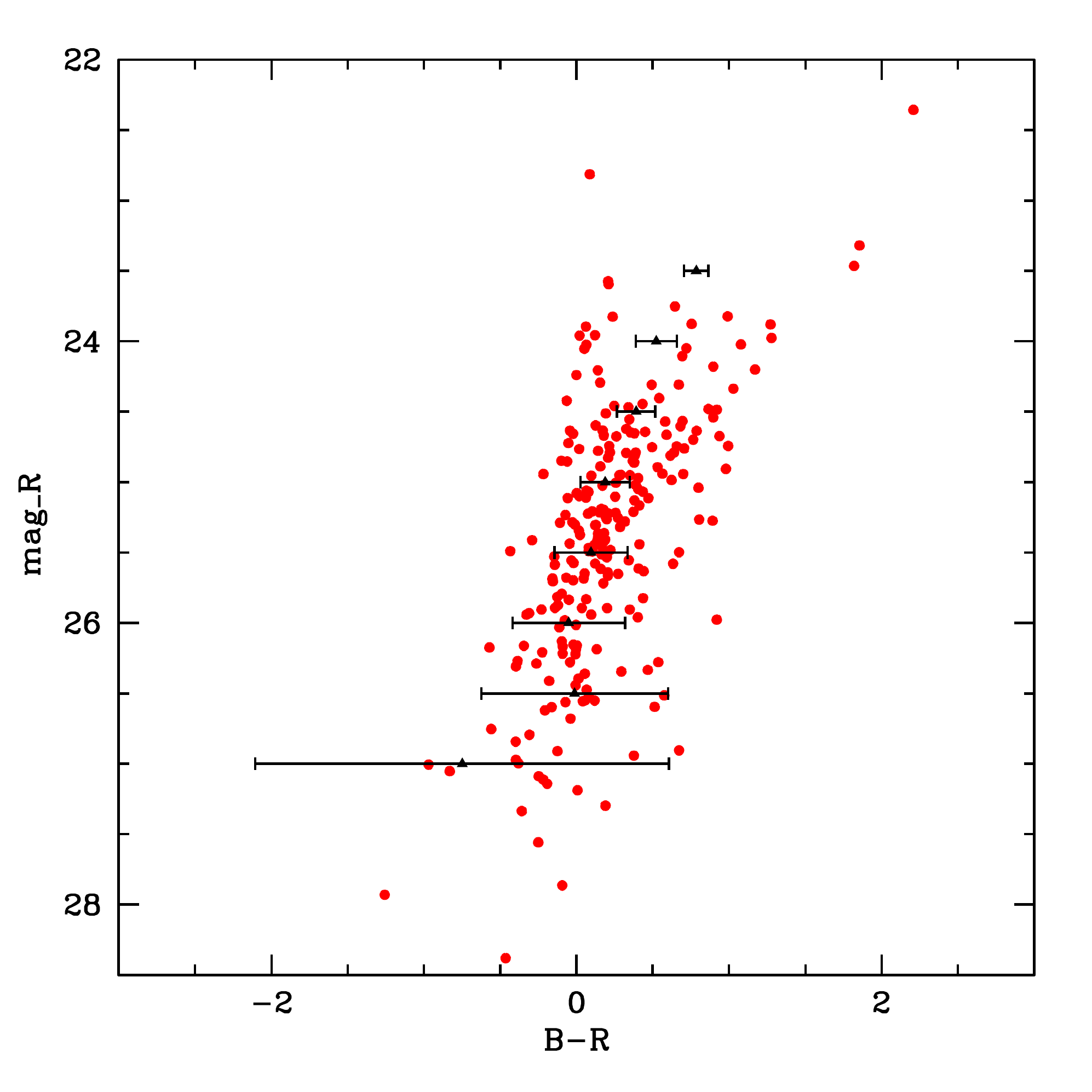}
\includegraphics[width=70mm]{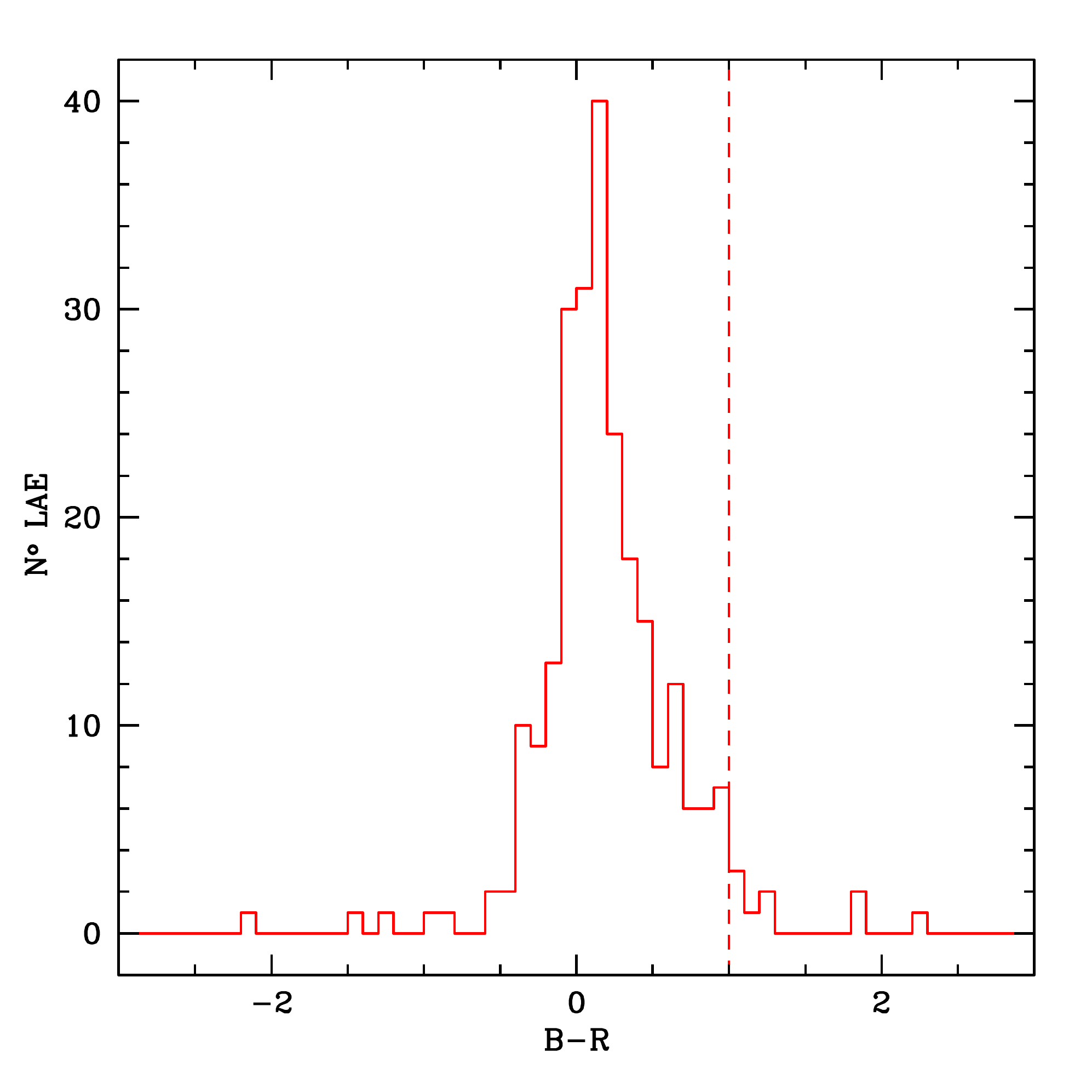}
\caption{{\it Left panel}: Broad-band $R$ magnitude as function of $B-R$ color. 
The median value of $B-R$ and median photometric uncertainty in each 0.5 $R$ magnitude
bin are shown by triangles and error bars respectively. For $R\geq27$ the median $B-R=-0.9\pm1.4$ and it is represented by the lower triangle and its error bar. {\it Right panel}: $B-R$ color distribution. The vertical dashed line represents $B-R=1$ below which there are the colors of the majority of the LAEs in our sample.} 
\label{fig:BR}
\end{center}
\end{figure}

\begin{figure}[htbp]
\begin{center}
\includegraphics[width=70mm]{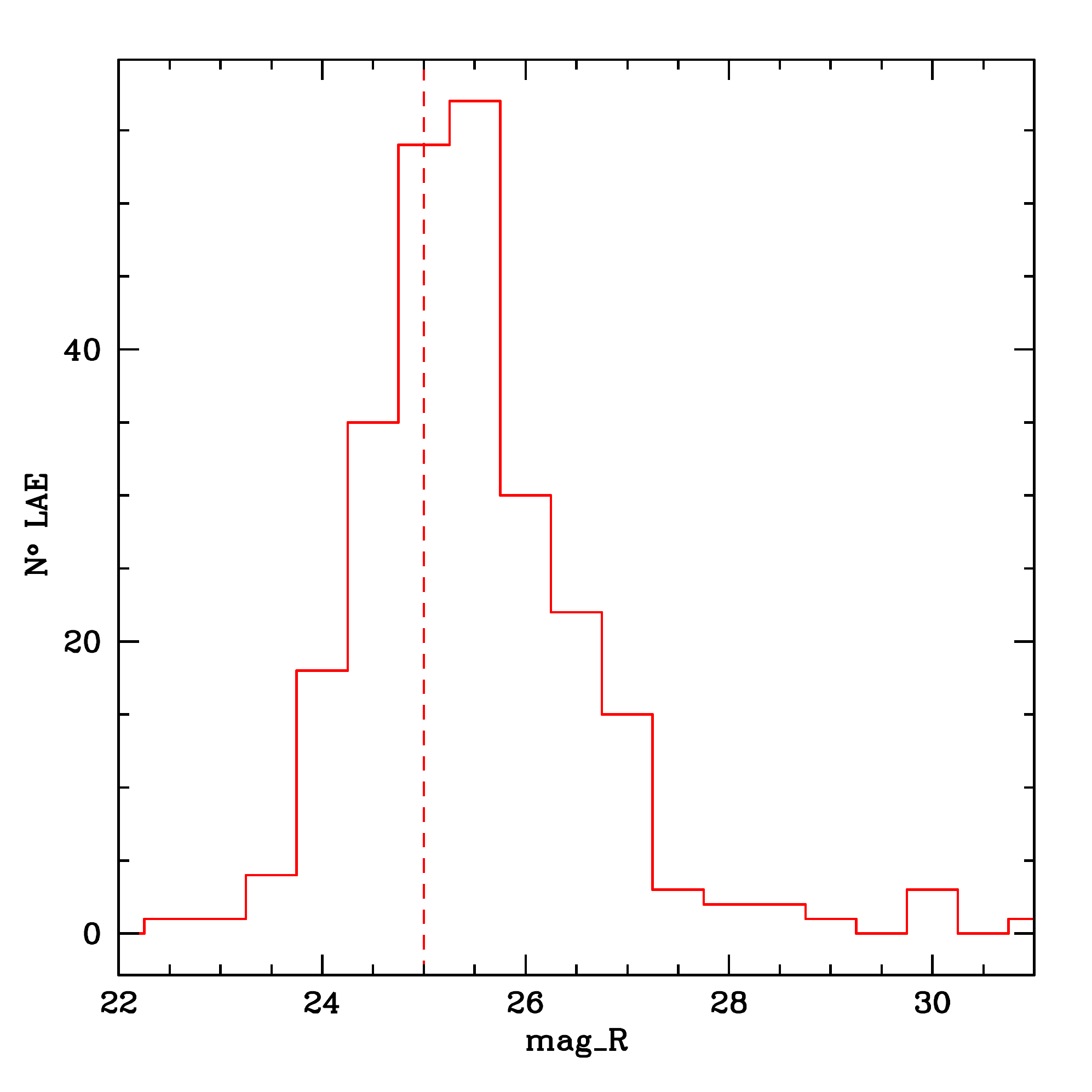}
\includegraphics[width=70mm]{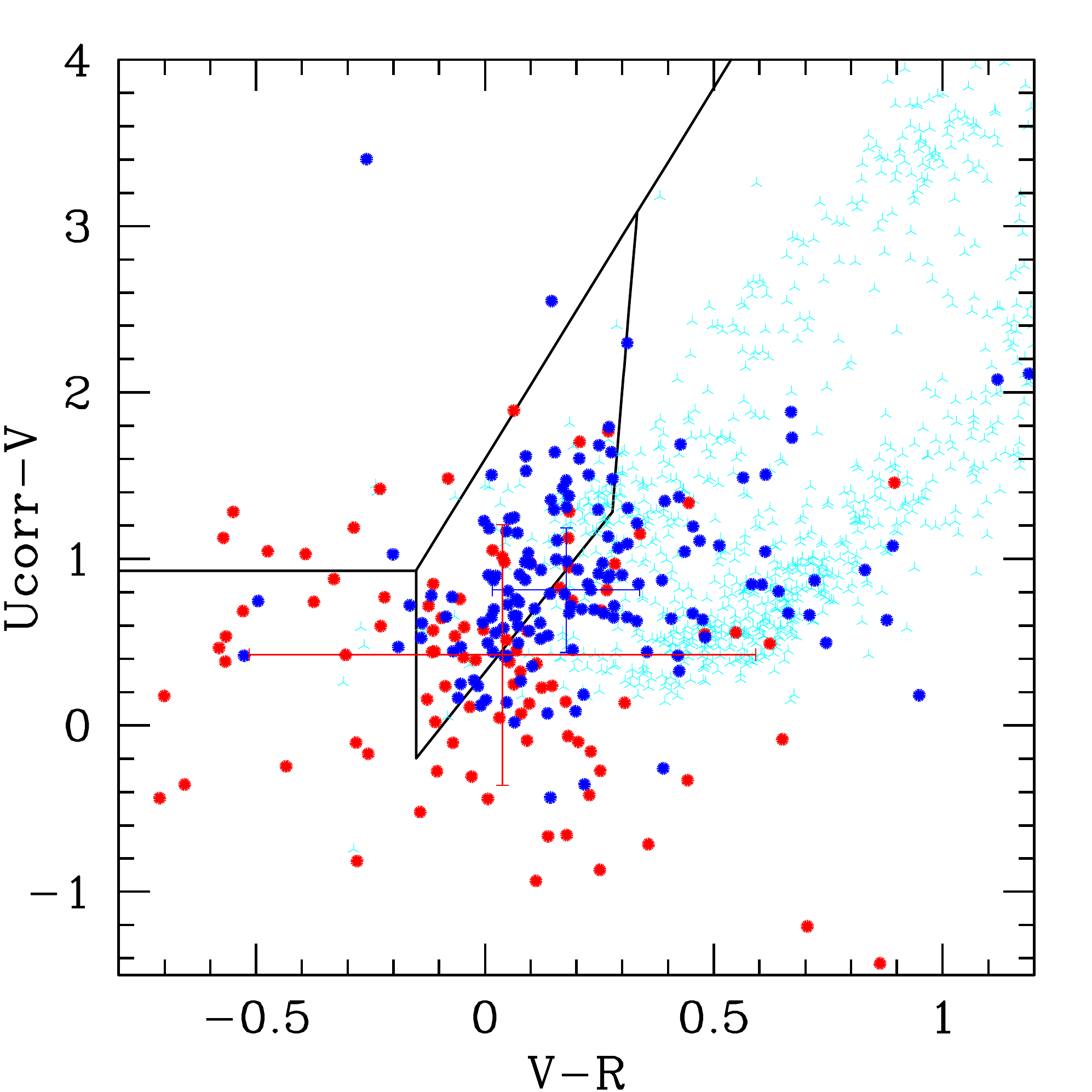}
\caption{{\it Left panel}:  $R$ magnitude distribution of the LAE sample. The vertical dashed line is drawn at $R=25$.  {\it Right panel}: Two-color diagram for $R<25$ (blue) and $R>25$ (red). Red (blue) error bars show median $Ucorr-V$ and $V-R$  colors and uncertainties of the faint (bright) subsamples. Cyan points are $z<1.4$ MUSYC spectroscopic catalog objects.}
\label{fig:uvr}
\end{center}
\end{figure}

\begin{figure}[htbp]
\begin{center}
\includegraphics[width=85mm]{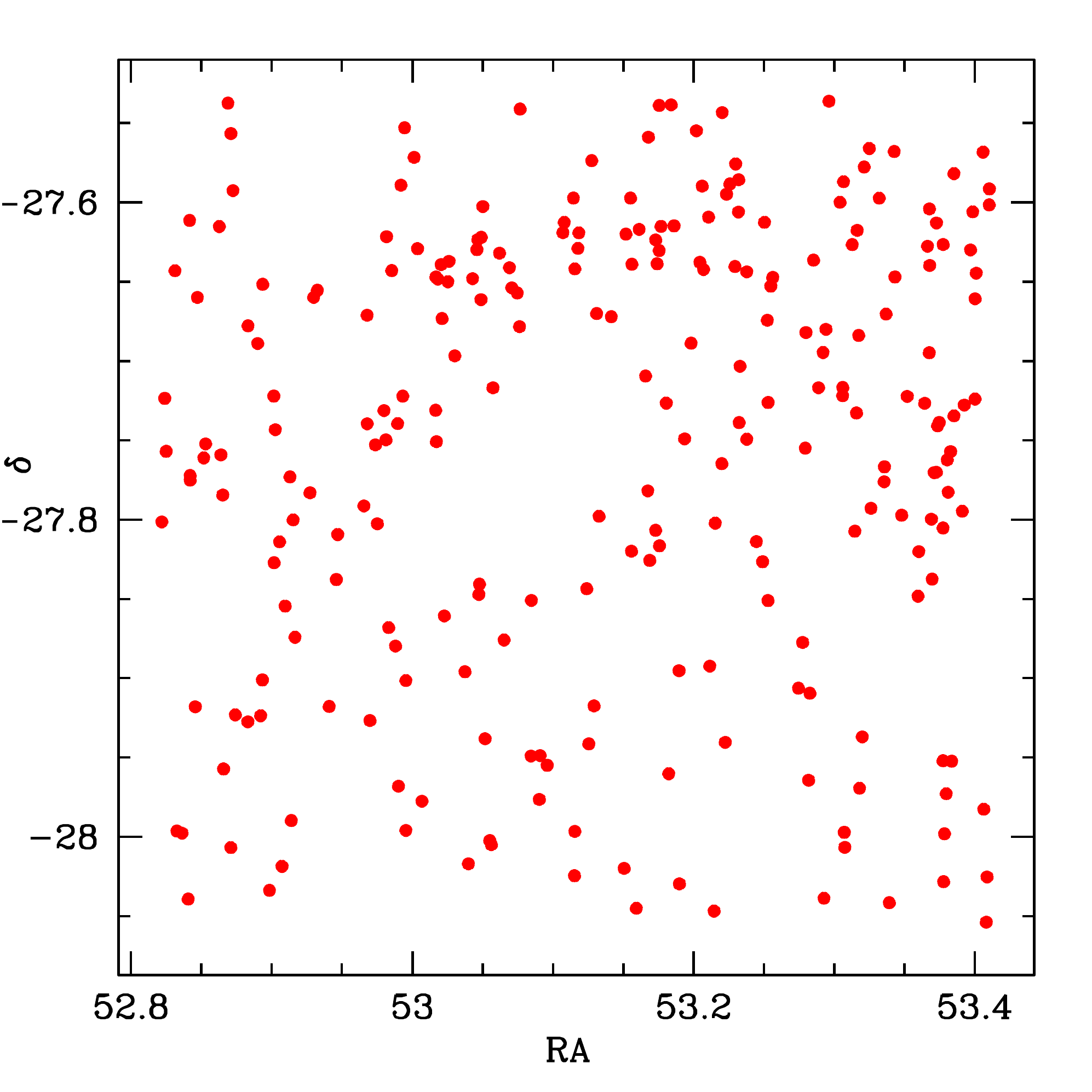}
\includegraphics[width=80mm]{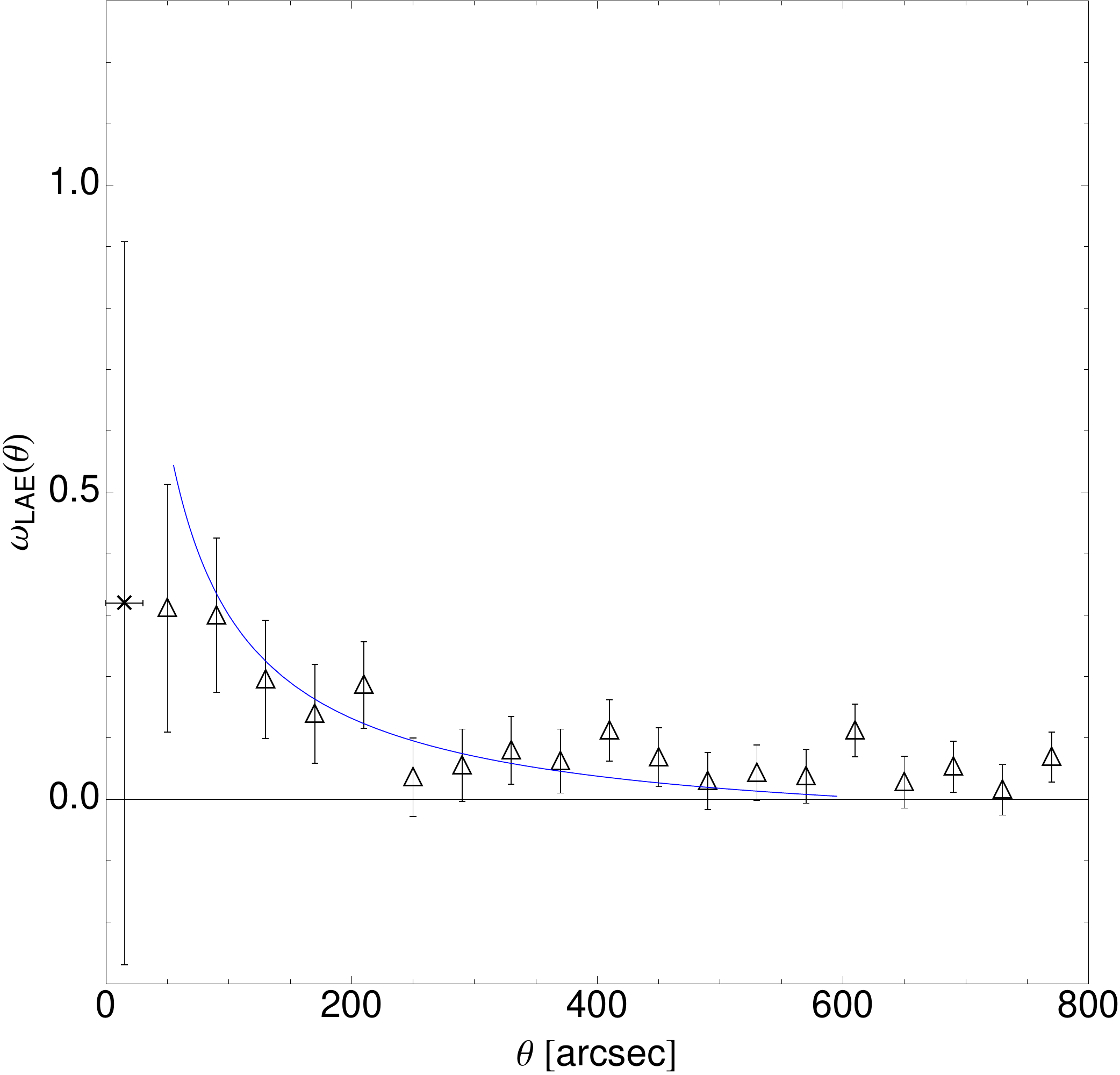}
\caption{ 
{\it Upper panel}: Declination versus right ascension plot, showing the spatial distribution of the 250 LAEs. 
{\it Lower panel}: Angular correlation function generated from our sample of 250 LAEs in black triangles. The blue solid curve is the best power law fit to the data, calculated from 40 to 600 arcsec. This range is chosen to avoid the 1-halo term at small scales and to avoid sampling noise at big scales.
} 
\label{fig:cluster}
\end{center}
\end{figure}

\begin{figure}[htbp]
\begin{center}
\includegraphics[width=80mm]{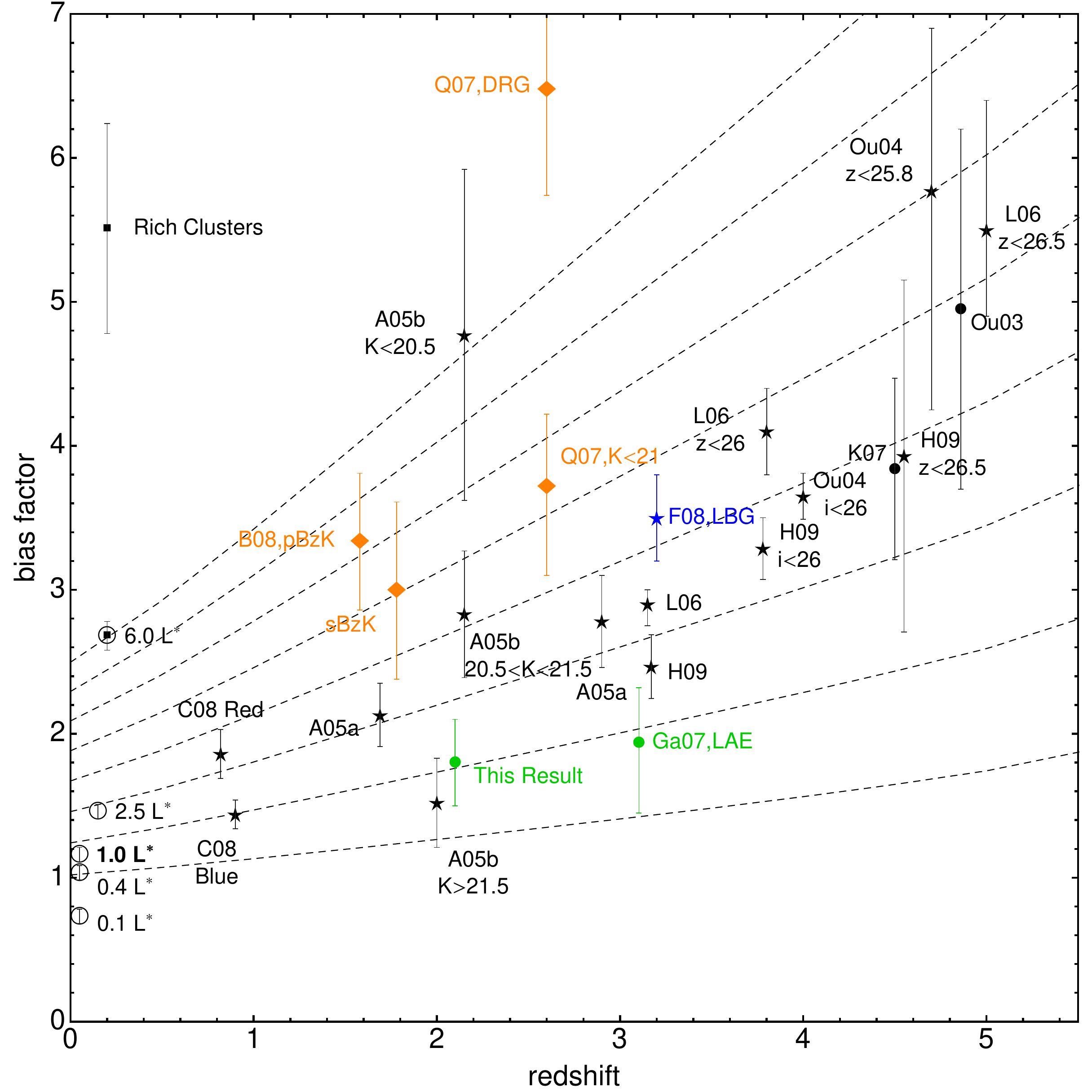}
\caption{ 
Bias evolution in redshift for a variety of
 galaxy populations. Stars are galaxies selected from their UV
 colors/magnitudes, diamonds are K selected, filled circles are
 narrow-band selected and open circles are low-redshift samples.
 The result from this work at $z\simeq2.1$ is plotted in green assuming a contamination fraction of 7$\pm$7\%, together with the
 MUSYC LAE result from \citeauthor{Gawiser:2007} (2007) at $z\simeq3.1$ (G07). Other results
 from MUSYC are also in color, namely: LBGs at $z\sim3$ in blue (F08)
 from \citeauthor{Francke:2008} (2008),  K-selected galaxies and
 Distant Red Galaxies (DRG) at $z\simeq 2.6$ from \citeauthor{Quadri:2007} (2007)
 (Q07), and both passive and star-forming, BzK-selected galaxies from
 \citeauthor{Blanc:2008} (2008) (B08), in orange.  The labels for
 other literature results correspond to: $z\simeq1.7$ color selected
 galaxies and LBG (A05a) from \citeauthor{Adelbergera:2005} (2005a), $z\simeq2.1$ color
 selected galaxies divided by their K magnitude (A05b) from
 \citeauthor{Adelbergerb:2005} (2005b), LBGs at $z\simeq3$, $z\simeq4$, $z\simeq5$ (L06, H09)
 from \citeauthor{Lee:2006} (2006) and from \citeauthor{Hild2009} (2009), LBGs at $z \simeq$~ 4 and 5 (Ou04) from \citeauthor{Ouchi:2004} (2004), high-z LAE measurements (Ou03, K07) from \citeauthor{Ouchi:2003} (2003)
 and \citeauthor{kovac:2007} (2007). At intermediate redshift, $z\sim1$,  galaxies
 separated by color (C08) come from \citeauthor{Coil:2008} (2008). Low redshift
 ($z\sim0$) galaxies, labeled by
 their optical luminosity, come from \citeauthor{Zehavi:2005} (2005) and the
 single point for rich galaxy clusters from \citeauthor{Banchall:2003} (2003). The dashed line corresponds to conditional mass function trajectories for bias evolution from Sheth \& Tormen theory.  
}
\label{fig:cluster2}
\end{center}
\end{figure}

We derived the pure continuum flux and the pure emission line flux from the observed fluxes in NB3727 and in the combination of $U$ and $B$ broad bands. \\ We model the LAE spectrum as an intrinsically constant continuum in frequency ($C_{\nu}$) plus a delta-function emission line in which the intergalactic medium (IGM) absorption is assumed negligible, i.e. 
\begin{equation}
f_{\nu, EL}(\lambda) = F_{EL} \frac{\lambda_{EL}^2}{c}\delta(\lambda_{EL}),
\label{a}
\end{equation}
\begin{equation}
f_{\nu, c}(\lambda) = e^{-\tau_{eff}(\lambda)} C_{\nu},
\label{aa}
\end{equation}
where
 $F_{EL}$ is the integrated flux inside the line in erg cm$^{-2}$ sec$^{-1}$, equal to $EW_{obs} \cdot f_{\lambda,c}(\lambda_{EL})$ and $C_{\nu}$ is the continuum flux density in erg cm$^{-2}$ sec$^{-1}$ Hz$^{-1}$.
Both emission line and continuum flux contribute to the NB3727 filter (NB) as:
\begin{equation}
f_{\nu,NB}=\frac{\int f_{\nu,EL}(\lambda) (c/\lambda^2) T_{NB}(\lambda) d\lambda}{\int (c/\lambda^2) T_{NB}(\lambda) d\lambda} + 
\frac{\int f_{\nu,c}(\lambda) (c/\lambda^2) T_{NB}(\lambda) d\lambda}{\int (c/\lambda^2) T_{NB}(\lambda) d\lambda} = 
\label{bb}
\end{equation}
\begin{equation}
\frac{   \int F_{EL} \frac {\lambda_{EL}^2}{c}\delta(\lambda_{EL}) (c/\lambda^2) T_{NB}(\lambda) d\lambda}  {\int (c/\lambda^2) T_{NB}(\lambda) d\lambda} + 
\frac{\int e^{-\tau_{eff}(\lambda)} C_{\nu} (c/\lambda^2) T_{NB}(\lambda) d\lambda}    {\int (c/\lambda^2) T_{NB}(\lambda) d\lambda} 
= \frac{F_{EL}T_{EL}}{\int (c/\lambda^2)T_{NB}(\lambda) d\lambda} + Q_{NB}C_{\nu}
\label{b}
\end{equation}
where the factor Q$_{NB}$ (\citealt{Venemans:2005}) is here defined as: $ \frac{\int e^{-\tau_{eff}(\lambda)} (c / \lambda^2) T_{NB}(\lambda) d\lambda }  {\int (c / \lambda^2) T_{NB}(\lambda) d\lambda} = 0.91$ and represents the fraction of the continuum that is transmitted after absorption by neutral hydrogen, averaged over the $NB3727$ bandpass.  T$_{EL}$=T$_{NB}(\lambda_{EL}$) has expectation value $<T_{EL}>_{PDF} = 0.124$, obtained convolving the filter transmission with a probability redshift distribution function (PDF) like that observed at $z\simeq3.1$ and taking the average. 
This way we use a filter transmission that best represents a typical Ly$\alpha$ emission line galaxy. If we used the maximum transmission of the filter, we would underestimate the Ly$\alpha$ fluxes.\\
We assume that inside the $B$ broad-band filter only the continuum is observed:
\begin{equation}
f_{\nu,B} =  \frac   {\int f_{\nu,c}(\lambda) (c/\lambda^2) T_{B}(\lambda) d\lambda}    {\int (c/\lambda^2) T_{B}(\lambda) d\lambda}   = \frac{\int e^{-\tau_{eff}(\lambda)} C_{\nu} (c/\lambda^2) T_{B}(\lambda) d\lambda}    {\int (c/\lambda^2) T_{B}(\lambda) d\lambda} = Q_{B} C_{\nu},
\label{c}
\end{equation}
where Q$_{B}$ is defined as: $ \frac{\int e^{-\tau_{eff}(\lambda)} (c / \lambda^2) T_{B}(\lambda) d\lambda }  {\int (c / \lambda^2) T_{B}(\lambda) d\lambda} = 0.999 \sim 1$,
but $U$ is just like NB3727 with emission line and continuum contributions:
\begin{equation}
f_{\nu,U} = \frac{\int f_{\nu,EL}(\lambda) (c/\lambda^2) T_{U}(\lambda) d\lambda}{\int (c/\lambda^2) T_{U}(\lambda) d\lambda} + 
\frac   {\int f_{\nu,c}(\lambda) (c/\lambda^2) T_{U}(\lambda) d\lambda}    {\int (c/\lambda^2) T_{U}(\lambda) d\lambda} = \frac{F_{EL}T_{EL,U}}{\int (c/\lambda^2)T_{U}(\lambda) d\lambda} + Q_{U}C_{\nu},
\label{d}
\end{equation}
also T$_{EL,U}$=T$_{U}(\lambda_{EL}$) can be calculated as the average in the PDF,  $<T_U>_{PDF} = 0.185$ and $Q_{U}$ is defined as: $ \frac{\int e^{-\tau_{eff}(\lambda)} (c / \lambda^2) T_{U}(\lambda) d\lambda }  {\int (c / \lambda^2) T_{U}(\lambda) d\lambda} = 0.89$.\\
This way we have:
\begin{equation}
f_{\nu,NB} = \frac{F_{EL} T_{EL}}{\int (c/\lambda^2) T_{NB}(\lambda) d\lambda} +  Q_{NB}C_{\nu} 
\label{f}
\end{equation}
\begin{equation}
f_{\nu,UB}= 0.8\frac{F_{EL} T_{EL,U}}{ \int (c/\lambda^2) T_{U}(\lambda) d\lambda} + Q_{UB} C_{\nu}.
\label{g}
\end{equation}
where $f_{\nu,UB} = 0.2 \cdot f_{\nu,B} + 0.8 \cdot f_{\nu,U}$ and Q$_{UB}$ is defined as: $0.8Q_{U} + 0.2 Q_{B} = 0.91$.\\
From this system of equations we derive $F_{EL}$ and $C_{\nu}$ as:
\begin{equation}
C_{\nu} = \frac{f_{\nu,UB} - c1f_{\nu, NB}} {Q_{UB} - c1 Q_{NB}}
\label{h}
\end{equation}
\begin{equation}
F_{EL} = (f_{\nu,NB} - Q_{NB} C _{\nu}) \int (c/\lambda^2) T_{NB}(\lambda) d\lambda /T_{EL}
\label{i}
\end{equation}
where c1 is constant depending on the filter shapes:
\begin{equation}
c1= \frac{T_{EL,U} 0.8 \int (c/\lambda^2) T_{NB}(\lambda) d\lambda} {T_{EL} \int{(c/\lambda^2) T_{U}(\lambda) d\lambda} } = 0.064,
\label{l}
\end{equation}
where the f$_{\nu}$s are the observed flux densities, estimated in $\mu$Jy by us.
We estimated the continuum subtracted emission line flux density as $f_{\nu,NB} - Q_{NB}C_{\nu} = 1.07 (f_{\nu,NB} - f_{\nu,UB})$. 
To define the broad-band color $U-V$, we need to subtract from the $U$ band the contribution of the emission line in the $U$ transmission filter as:\\
$f_{\nu,U,only~ continuum}= f_{\nu, U} - f_{\nu,U} ^{EL} = f_{\nu,U} - \frac{F_{EL} T_{EL,U}  }{ \int (c/\lambda^2) T_{U}(\lambda) d\lambda}.$\\
The $UB-NB3727$ color is calculated from the ratio of observed fluxes $f_{\nu,NB}$ and $f_{\nu,UB}$.
As we introduced before $F_{EL}/ f_{\lambda,c}(\lambda_{EL}) = EW_{obs}$, so we replaced $F_{EL} = EW_{obs} C_{\nu} (c/ \lambda_{EL}^2)$, taking into account the IGM absorption. 
So:
\begin{equation}
f_{\nu,NB}=C_{\nu} Q_{NB} + \frac{EW_{obs} C_{\nu} (c/ \lambda_{EL}^2) T_{EL}}{\int (c/\lambda^2) T_{NB}(\lambda)d\lambda}
\label{u}
\end{equation}
and
\begin{equation}
f_{\nu,UB}= C_{\nu} Q_{UB}+ \frac{0.8 ~ EW_{obs} C_{\nu} (c/ \lambda_{EL}^2) T_{EL,U}}{\int (c/\lambda^2) T_{U}(\lambda)d\lambda}
\label{v}
\end{equation}
For EW$_{obs} = 20 \times$ (1+z) = 61 {\AA}, hence $\frac{f_{\nu,NB}}{f_{\nu,UB}} = 1.97$.\\
Therefore:
\begin{equation}
 UB-\mathrm{NB3727} = 2.5 ~ log (\frac{f_{\nu,NB}}{f_{\nu,UB}}) > 0.73
 \label{colorequation}
 \end{equation}
is the color cut we are using as the first selection criterion of LAEs to select galaxies with rest-frame EW bigger than 20 {\AA}. 
Plugging equations (\ref{u}), (\ref{v}) into the equality of (\ref{colorequation}), we derive an expression for 
\begin{equation}
EW_{obs}=A/B, 
\label{ewequation}
\end{equation}
where
\begin{equation}
A=Q_{NB} - Q_{UB} 10^{((UB-NB3727)/2.5)}
\label{A}
\end{equation}
and 
\begin{equation}
B=\frac{0.8 T_{EL,U} (c/\lambda_{EL}^2) 10^{((UB-NB3727)/2.5)}} {\int(c/\lambda^2)T_{U}(\lambda)d\lambda}-\frac{T_{EL}(c/\lambda_{EL}^2)}{\int(c/\lambda^2)T_{NB}(\lambda)d\lambda}
\label{B}
\end{equation}
In the case in which the $EW_{obs}$ is infinite, for the pure emission line objects, the parts of the expressions containing EW dominate and the $EW_{obs}$ simplifies, giving a maximum theoretical value of $UB$-NB3727~$=2.5 ~log (15.42)=2.97$. 
Spurious objects that appear only in the narrow-band should have infinite $UB$-NB3727, but may scatter below the maximum theoretical value due to photometric errors.  Real LAEs often have faint continuum, so photometric errors can scatter their $UB$-NB3727 color above this theoretical maximum.  Hence the maximum cannot be used as a sharp discriminator between real and fake objects, and it is inevitable that a few objects will appear to have $EW_{obs} = \infty$.
Among our LAEs, 4 of them have these ``unphysical'' values of $UB-$NB3727, consistent with our upper limit on spurious objects found above, but we do not know which of these objects are truly spurious or simply fell prey to negative noise fluctuations in their $UB$ continua. In Fig.~\ref{fig:ew} we associated the value EW$=$400 {\AA} to these 4 objects.

\clearpage

\clearpage

\clearpage

\input{tab0.tex}
\input{tab1.tex}
\input{tab2.tex}

\input{tab3.tex}


\end{document}

%% file: tab0.tex
\begin{deluxetable}{lccr}
\tablewidth{0pt}
\tablecaption{LOG OF ALL BANDS}
\tablehead{
\colhead{Band}           & \colhead{Telescope}      &
\colhead{Exposure Time(s)}          & \colhead{5$\sigma$ detection limit}  }
\startdata
NB3727 & 4m CTIO   & 128700  & 25.1\\
$U$ & 2.2m La Silla & 78891 & 26.1\\
$B$ & 2.2m La Silla & 69431 & 26.9\\
$V$ & 2.2m La Silla & 104603 & 26.5\\
$R$ & 2.2m La Silla & 87653 & 26.5\\
\enddata
\label{tab:allbands}
\end{deluxetable}

%% file: tab1.tex
\begin{deluxetable}{lccr}
\tablewidth{0pt}
\tablecaption{LOG OF NARROW BAND OBSERVATIONS}
\tablehead{
\colhead{(UT) starting date}           & \colhead{NB Exposure (s)}      &
\colhead{Seeing FWHM ($''$)}          & \colhead{Conditions}  }
\startdata
2007 Dec., 3rd & 10800   &   1.4 & Full nt, cloudless\\
2007 Dec., 4rd & 7200 & 1.5 & Half nt, cloudless\\
2007 Dec., 5th & 21600 & 1.3 & Full nt, cloudless\\
2007 Dec., 6th & 10800 & 1.3 & Half nt, cloudless\\
2007 Dec., 7th & 21600 & 1.4 & Full nt, cloudless\\
2007 Dec., 8th & 10800 & 1.7 & Half nt, cloudless\\
2007 Dec., 9th & 18000 & 1.6 & Full nt, cloudless\\
2007 Dec., 10th & 3600 & 1.1 & Half nt, cloudless\\
2007 Dec., 11th & 14400 & 1.2 & Full nt, cloudless\\
2007 Dec., 12th & 9900 & 1.3 & Half nt, cloudless\\
\enddata
\tablecomments{The seeing here reported is the mean of the mode of the seeing estimations for each night.\\Full nt = full night allocated\\Half nt = second half of the night allocated}
\label{tab:obs}
\end{deluxetable}

%% file: tab2.tex
\begin{deluxetable}{lccccr}
\tablewidth{0pt}
\tablecaption{LOG OF NARROW BAND IMAGES PROPERTIES}
\tablehead{
\colhead{image}           & \colhead{airmass}      &
\colhead{MSCSCALE}          & \colhead{skyRMS}  &
\colhead{seeing ($''$)}          & 
\colhead{weight}}
\startdata
3dec\_1 &  1.002 &  1.000   &    7.97 &1.35 &    0.00858 \\
 3dec\_2 &  1.042 &  1.037   &    8.11 &  1.26    &   0.00871 \\
 3dec\_3  & 1.154 & 1.128     &  8.68  &1.63    &  0.00376 \\
 4dec\_1  & 1.066 & 1.215     &  7.20 & 1.59   &    0.00500  \\
 4dec\_2  & 1.203 & 1.305     &  7.44 & 1.46    &   0.00493 \\
 5dec\_1  & 1.091 & 1.199     &  6.16 & 1.65    &  0.00642   \\
 5dec\_2  & 1.016 & 1.149     &  6.34 & 1.23    &   0.01207 \\
 5dec\_3  &  1.003 & 1.168     &  6.74 & 1.18     &   0.01104 \\
 5dec\_4  &  1.051 & 1.203     &  6.84 & 1.16     &   0.01034  \\
 5dec\_5  & 1.172 & 1.287     &  7.21  & 1.29     &   0.00687  \\
 5dec\_6 & 1.411  & 1.470     &  7.71  &1.39   &    0.00400 \\
 6dec\_1  & 1.072 & 1.178      & 7.14 & 1.42      &  0.00697   \\
 6dec\_2  & 1.218 & 1.257       & 7.59 & 1.30  &   0.00641 \\
 6dec\_3  & 1.504 & 1.472  &     8.27 & 1.34     &   0.00373  \\
 7dec\_1  & 1.104 & 1.055    &   8.16 & 1.62    &   0.00495  \\
 7dec\_2  &  1.019 & 1.044    &   8.24 & 1.44    &   0.00647 \\
 7dec\_3  & 1.003 & 1.056    &   8.64 & 1.41  &   0.00600  \\
 7dec\_4  &  1.049 & 1.073    &   8.92 & 1.32    &  0.00619 \\  
 7dec\_5 &  1.168 &  1.000  &    8.87 &  1.14    &   0.00979 \\ 
 7dec\_6   &  1.404 &1.135    &  9.66 & 1.22  &   0.00589  \\
 8dec\_1  &  1.099 & 1.062    &  7.95 & 1.44   &    0.00752  \\
 8dec\_2 &  1.268  & 1.184   &   8.15 & 1.60   &    0.00463  \\
 8dec\_3  &  1.599 & 1.505   &   8.76 & 2.06    &    0.00130  \\
 9dec\_1  &    1.093 & 1.00   &   7.51 & 1.60   &    0.00761  \\
 9dec\_2  &  1.016 & 0.987   &   7.23 & 1.57      &  0.00881  \\
 9dec\_3  &  1.003 & 0.994   &   7.17 & 1.63   &   0.00814  \\
 9dec\_4  &  1.003 & 1.014   &   7.66 & 1.38  &    0.00961 \\
 9dec\_5  &  1.169 & 1.092  &   8.16 & 1.64   &   0.00512 \\
 10dec\_1  &  1.435 & 1.184   &   8.29 & 1.14   &   0.00800 \\
 11dec\_1  &   1.001 & 0.969    &  6.40 & 0.99  &   0.02291  \\
 11dec\_2  &  1.036 & 1.003   &   6.55 & 0.97   &   0.02072   \\
 11dec\_3  &   1.143 & 1.067   &   6.65 & 1.30     &   0.01279 \\
 11dec\_4  &   1.356 & 1.208   &   6.79 & 1.53  &    0.00706 \\
 12dec\_1  &  1.142 & 1.073   &   7.40& 0.99    &   0.01397  \\
 12dec\_2 &   1.355 &1.216   &   7.84 & 1.21   &     0.00786 \\
 12dec\_3  &  1.775 & 1.989    &  6.98  &1.57  &  0.00233  \\
\enddata
\label{tab:images}
\end{deluxetable}

%% file: tab3.tex
\begin{deluxetable}{cccccccccc}
\tabletypesize{\scriptsize}
\rotate
\tablecaption{Properties of LAEs binned by continuum magnitude}
\tablewidth{0pt}
\tablehead{
\colhead{Rmin}      &   \colhead{Rmax}      &
\colhead{number of objects}        &
\colhead{B-R}          &      \colhead{Ucorr-V}    &
\colhead{V-R}   &    {UB-NB3727}     &      \colhead{$EW_{rest frame}$}    &
\colhead{SFR(Ly$\alpha$)}          &      \colhead{SFR(UV)}}
\startdata
 - &  24.0    &  15 &  0.6$\pm$0.7  &  1.1$\pm$1.0   &     0.1$\pm$0.5 &  1.0$\pm$0.2 &  30$\pm$13  &   1.5$\pm$7.4   &      15.6$\pm$12.3  \\
24.0 &  24.5  & 20 &  0.5$\pm$0.4 &  1.2$\pm$0.6  &      0.3$\pm$0.2 &    0.8$\pm$0.2 &   25$\pm$11  &    1.0$\pm$1.5  &     10.0$\pm$1.5 \\
24.5 &   25.0 & 51  &  0.4$\pm$0.3  &   0.9$\pm$0.4  &     0.2$\pm$0.2 &   0.9$\pm$0.2  &  28$\pm$11 &   1.0$\pm$0.6   &    6.6$\pm$0.8   \\
25.0 &  25.5  & 62 & 0.2$\pm$0.2  &  0.7$\pm$0.5  &     0.1$\pm$0.2 &    1.0$\pm$0.3&   32$\pm$30  &    1.0$\pm$0.9  &     4.1$\pm$0.6 \\
25.5 & 26.0 & 39  &   0.1$\pm$0.3  &  0.6$\pm$0.5   &    0.0$\pm$0.2 &   1.2$\pm$0.4  &  43$\pm$48   &   1.1$\pm$0.7  &     2.8$\pm$0.4 \\
26.0 &  26.5 & 25 &  -0.0$\pm$0.2 &  0.5$\pm$0.7  &     0.1$\pm$0.4&    1.7$\pm$0.3 &  96$\pm$51 &  1.4$\pm$1.1   &    1.7$\pm$0.2  \\
26.5 &  27.0  & 19  &   -0.0$\pm$0.8  &   -0.1$\pm$0.7   &    0.1$\pm$0.4  &   1.7$\pm$0.4  &  101$\pm$130  &  1.5$\pm$0.4   &   1.1$\pm$0.2 \\
27.0 &  -  & 19 &  -0.3$\pm$1.4 &   -0.2$\pm$1.5  &    -0.6$\pm$1.7 &    2.0$\pm$1.1  &   150$\pm$456  &  1.5$\pm$0.2  &     0.4$\pm$0.3  \\
\enddata
\tablecomments{The R, B-R, Ucorr-V, V-R, UB-NB3727, EW ({\AA}) and SFRs (M$_{\odot}$ yr$^{-1}$) are the median in the R magnitude bins in the range between Rmin and Rmax. The errors shown are the standard deviation in each bin.}
\label{tab:Rchar}
\end{deluxetable}

%% file: ms.bbl
\clearpage


\begin{thebibliography}{62}
\expandafter\ifx\csname natexlab\endcsname\relax\def\natexlab#1{#1}\fi

\bibitem[{{Adelberger} {et~al.}(2005{\natexlab{a}}){Adelberger}, {Steidel},
  {Pettini}, {Shapley}, {Reddy}, \& {Erb}}]{Adelbergera:2005}
{Adelberger}, K.~L., {Steidel}, C.~C., {Pettini}, M., {Shapley}, A.~E.,
  {Reddy}, N.~A., \& {Erb}, D.~K. 2005{\natexlab{a}}, \apj, 619, 697
  
\bibitem[{{Adelberger} {et~al.}(2005{\natexlab{b}}){Adelberger}, {Erb},
  {Steidel}, {Reddy}, {Pettini}, \& {Shapley}}]{Adelbergerb:2005}
{Adelberger}, K.~L., {Erb}, D.~K., {Steidel}, C.~C., {Reddy}, N.~A., {Pettini},
  M., \& {Shapley}, A.~E. 2005{\natexlab{b}}, \apjl, 620, L75

\bibitem[{{Atek} {et~al.}(2009){Atek}, {Kunth}, {Schaerer}, {Hayes},
  {Deharveng}, {{\"O}stlin}, \& {Mas-Hesse}}]{Atek:2009}
{Atek}, H., {Kunth}, D., {Schaerer}, D., {Hayes}, M., {Deharveng}, J.~M.,
  {{\"O}stlin}, G., \& {Mas-Hesse}, J.~M. 2009, \aap, 506, L1

\bibitem[{{Bahcall} {et~al.}(2003){Bahcall}, {Dong}, {Hao}, {Bode}, {Annis},
  {Gunn}, \& {Schneider}}]{Banchall:2003}
{Bahcall}, N.~A., {Dong}, F., {Hao}, L., {Bode}, P., {Annis}, J., {Gunn},
  J.~E., \& {Schneider}, D.~P. 2003, \apj, 599, 814

\bibitem[{{Bertin} \& {Arnouts}(1996)}]{bertin1996}
{Bertin}, E. \& {Arnouts}, S. 1996, \aaps, 117, 393

\bibitem[{{Blanc} {et~al.}(2008){Blanc}, {Lira}, {Barrientos}, {Aguirre},
  {Francke}, {Taylor}, {Quadri}, {Marchesini}, {Infante}, {Gawiser}, {Hall},
  {Willis}, {Herrera}, \& {Maza}}]{Blanc:2008}
{Blanc}, G.~A., {Lira}, P., {Barrientos}, L.~F., {Aguirre}, P., {Francke}, H.,
  {Taylor}, E.~N., {Quadri}, R., {Marchesini}, D., {Infante}, L., {Gawiser},
  E., {Hall}, P.~B., {Willis}, J.~P., {Herrera}, D., \& {Maza}, J. 2008, \apj,
  681, 1099

\bibitem[{{Bouwens} {et~al.}(2006){Bouwens}, {Illingworth}, {Blakeslee}, \&
  {Franx}}]{Bouwens:2006}
{Bouwens}, R.~J., {Illingworth}, G.~D., {Blakeslee}, J.~P., \& {Franx}, M.
  2006, \apj, 653, 53

\bibitem[{{Bruzual} \& {Charlot}(2003)}]{Bruzual:2003}
{Bruzual}, G. \& {Charlot}, S. 2003, \mnras, 344, 1000

\bibitem[{{Coil} {et~al.}(2008){Coil}, {Newman}, {Croton}, {Cooper}, {Davis},
  {Faber}, {Gerke}, {Koo}, {Padmanabhan}, {Wechsler}, \& {Weiner}}]{Coil:2008}
{Coil}, A.~L., {Newman}, J.~A., {Croton}, D., {Cooper}, M.~C., {Davis}, M.,
  {Faber}, S.~M., {Gerke}, B.~F., {Koo}, D.~C., {Padmanabhan}, N., {Wechsler},
  R.~H., \& {Weiner}, B.~J. 2008, \apj, 672, 153

\bibitem[{{Dunkley} {et~al.}(2009){Dunkley}, {Komatsu}, {Nolta}, {Spergel},
  {Larson}, {Hinshaw}, {Page}, {Bennett}, {Gold}, {Jarosik}, {Weiland},
  {Halpern}, {Hill}, {Kogut}, {Limon}, {Meyer}, {Tucker}, {Wollack}, \&
  {Wright}}]{Dunkley:2009}
{Dunkley}, J., {Komatsu}, E., {Nolta}, M.~R., {Spergel}, D.~N., {Larson}, D.,
  {Hinshaw}, G., {Page}, L., {Bennett}, C.~L., {Gold}, B., {Jarosik}, N.,
  {Weiland}, J.~L., {Halpern}, M., {Hill}, R.~S., {Kogut}, A., {Limon}, M.,
  {Meyer}, S.~S., {Tucker}, G.~S., {Wollack}, E., \& {Wright}, E.~L. 2009,
  \apjs, 180, 306

\bibitem[{{Finkelstein} {et~al.}(2008){Finkelstein}, {Rhoads}, {Malhotra},
  {Grogin}, \& {Wang}}]{Fin:2008}
{Finkelstein}, S.~L., {Rhoads}, J.~E., {Malhotra}, S., {Grogin}, N., \& {Wang},
  J. 2008, \apj, 678, 655
  
\bibitem[{{Finkelstein} {et~al.}(2009){Finkelstein}, {Rhoads}, {Malhotra}, \&
  {Grogin}}]{Fyougold:2009}
{Finkelstein}, S.~L., {Rhoads}, J.~E., {Malhotra}, S., \& {Grogin}, N. 2009,
  \apj, 691, 465  

\bibitem[{{Francke} {et~al.}(2008){Francke}, {Gawiser}, {Lira}, {Treister},
  {Virani}, {Cardamone}, {Urry}, {van Dokkum}, \& {Quadri}}]{Francke:2008}
{Francke}, H., {Gawiser}, E., {Lira}, P., {Treister}, E., {Virani}, S.,
  {Cardamone}, C., {Urry}, C.~M., {van Dokkum}, P., \& {Quadri}, R. 2008,
  \apjl, 673, L13
  
\bibitem[{{Gawiser} {et~al.}(2006{\natexlab{a}}){Gawiser}, {van Dokkum},
  {Herrera}, {Maza}, {Castander}, {Infante}, {Lira}, {Quadri}, {Toner},
  {Treister}, {Urry}, {Altmann}, {Assef}, {Christlein}, {Coppi}, {Dur{\'a}n},
  {Franx}, {Galaz}, {Huerta}, {Liu}, {L{\'o}pez}, {M{\'e}ndez}, {Moore},
  {Rubio}, {Ruiz}, {Toft}, \& {Yi}}]{Gawiser:2006a}
{Gawiser}, E., {van Dokkum}, P.~G., {Herrera}, D., {Maza}, J., {Castander},
  F.~J., {Infante}, L., {Lira}, P., {Quadri}, R., {Toner}, R., {Treister}, E.,
  {Urry}, C.~M., {Altmann}, M., {Assef}, R., {Christlein}, D., {Coppi}, P.~S.,
  {Dur{\'a}n}, M.~F., {Franx}, M., {Galaz}, G., {Huerta}, L., {Liu}, C.,
  {L{\'o}pez}, S., {M{\'e}ndez}, R., {Moore}, D.~C., {Rubio}, M., {Ruiz},
  M.~T., {Toft}, S., \& {Yi}, S.~K. 2006{\natexlab{a}}, \apjs, 162, 1  

\bibitem[{{Gawiser} {et~al.}(2006{\natexlab{b}}){Gawiser}, {van Dokkum},
  {Gronwall}, {Ciardullo}, {Blanc}, {Castander}, {Feldmeier}, {Francke},
  {Franx}, {Haberzettl}, {Herrera}, {Hickey}, {Infante}, {Lira}, {Maza},
  {Quadri}, {Richardson}, {Schawinski}, {Schirmer}, {Taylor}, {Treister},
  {Urry}, \& {Virani}}]{Gawiser:2006b}
{Gawiser}, E., {van Dokkum}, P.~G., {Gronwall}, C., {Ciardullo}, R., {Blanc},
  G.~A., {Castander}, F.~J., {Feldmeier}, J., {Francke}, H., {Franx}, M.,
  {Haberzettl}, L., {Herrera}, D., {Hickey}, T., {Infante}, L., {Lira}, P.,
  {Maza}, J., {Quadri}, R., {Richardson}, A., {Schawinski}, K., {Schirmer}, M.,
  {Taylor}, E.~N., {Treister}, E., {Urry}, C.~M., \& {Virani}, S.~N.
  2006{\natexlab{b}}, \apjl, 642, L13
  
\bibitem[{{Gawiser} {et~al.}(2007){Gawiser}, {Francke}, {Lai}, {Schawinski},
  {Gronwall}, {Ciardullo}, {Quadri}, {Orsi}, {Barrientos}, {Blanc}, {Fazio},
  {Feldmeier}, {Huang}, {Infante}, {Lira}, {Padilla}, {Taylor}, {Treister},
  {Urry}, {van Dokkum}, \& {Virani}}]{Gawiser:2007}
{Gawiser}, E., {Francke}, H., {Lai}, K., {Schawinski}, K., {Gronwall}, C.,
  {Ciardullo}, R., {Quadri}, R., {Orsi}, A., {Barrientos}, L.~F., {Blanc},
  G.~A., {Fazio}, G., {Feldmeier}, J.~J., {Huang}, J.-S., {Infante}, L.,
  {Lira}, P., {Padilla}, N., {Taylor}, E.~N., {Treister}, E., {Urry}, C.~M.,
  {van Dokkum}, P.~G., \& {Virani}, S.~N. 2007, \apj, 671, 278  

\bibitem[{{Giavalisco} {et~al.}(2004){Giavalisco}, {Dickinson}, {Ferguson},
  {Ravindranath}, {Kretchmer}, {Moustakas}, {Madau}, {Fall}, {Gardner},
  {Livio}, {Papovich}, {Renzini}, {Spinrad}, {Stern}, \& {Riess}}]{Giav2004}
{Giavalisco}, M., {Dickinson}, M., {Ferguson}, H.~C., {Ravindranath}, S.,
  {Kretchmer}, C., {Moustakas}, L.~A., {Madau}, P., {Fall}, S.~M., {Gardner},
  J.~P., {Livio}, M., {Papovich}, C., {Renzini}, A., {Spinrad}, H., {Stern},
  D., \& {Riess}, A. 2004, \apjl, 600, L103

\bibitem[{{Gronwall} {et~al.}(2007){Gronwall}, {Ciardullo}, {Hickey},
  {Gawiser}, {Feldmeier}, {van Dokkum}, {Urry}, {Herrera}, {Lehmer}, {Infante},
  {Orsi}, {Marchesini}, {Blanc}, {Francke}, {Lira}, \&
  {Treister}}]{Gronwall:2007}
{Gronwall}, C., {Ciardullo}, R., {Hickey}, T., {Gawiser}, E., {Feldmeier},
  J.~J., {van Dokkum}, P.~G., {Urry}, C.~M., {Herrera}, D., {Lehmer}, B.~D.,
  {Infante}, L., {Orsi}, A., {Marchesini}, D., {Blanc}, G.~A., {Francke}, H.,
  {Lira}, P., \& {Treister}, E. 2007, \apj, 667, 79

\bibitem[{{Guhathakurta} {et~al.}(1990){Guhathakurta}, {Tyson}, \&
  {Majewski}}]{Gu:1990}
{Guhathakurta}, P., {Tyson}, J.~A., \& {Majewski}, S.~R. 1990, \apjl, 357, L9

\bibitem[{{Hildebrandt} {et~al.}(2006){Hildebrandt}, {Erben}, {Dietrich},
  {Cordes}, {Haberzettl}, {Hetterscheidt}, {Schirmer}, {Schmithuesen},
  {Schneider}, {Simon}, \& {Trachternach}}]{Hildebrandt:2006}
{Hildebrandt}, H., {Erben}, T., {Dietrich}, J.~P., {Cordes}, O., {Haberzettl},
  L., {Hetterscheidt}, M., {Schirmer}, M., {Schmithuesen}, O., {Schneider}, P.,
  {Simon}, P., \& {Trachternach}, C. 2006, \aap, 452, 1121
  
\bibitem[{{Hildebrandt} {et~al.}(2009){Hildebrandt}, {Pielorz}, {Erben},
  {van Waerbeke}, {Simon}, \& {Capak}}]{Hild2009}
{Hildebrandt}, H., {Pielorz}, J., {Erben}, T.., {van Waerbeke}, L.,
  {Simon}, P., \& {Capak}, P. 2009, \aap, 498, 725  
  
\bibitem[{{Hogg} {et~al.}(1998){Hogg}, {Cohen}, {Blandford}, \&
  {Pahre}}]{Hogg:1998}
{Hogg}, D.~W., {Cohen}, J.~G., {Blandford}, R., \& {Pahre}, M.~A. 1998, \apj,
  504, 622

\bibitem[{{Kova{\v c}} {et~al.}(2007){Kova{\v c}}, {Somerville}, {Rhoads},
  {Malhotra}, \& {Wang}}]{kovac:2007}
{Kova{\v c}}, K., {Somerville}, R.~S., {Rhoads}, J.~E., {Malhotra}, S., \&
  {Wang}, J. 2007, \apj, 668, 15

\bibitem[{{Lai} {et~al.}(2008){Lai}, {Huang}, {Fazio}, {Gawiser}, {Ciardullo},
  {Damen}, {Franx}, {Gronwall}, {Labbe}, {Magdis}, \& {van Dokkum}}]{Lai:2008}
{Lai}, K., {Huang}, J.-S., {Fazio}, G., {Gawiser}, E., {Ciardullo}, R.,
  {Damen}, M., {Franx}, M., {Gronwall}, C., {Labbe}, I., {Magdis}, G., \& {van
  Dokkum}, P. 2008, \apj, 674, 70

\bibitem[{{Landy} \& {Szalay}(1993)}]{Landy:1993}
{Landy}, S.~D. \& {Szalay}, A.~S. 1993, \apj, 412, 64

\bibitem[{{Lee} {et~al.}(2006){Lee}, {Giavalisco}, {Gnedin}, {Somerville},
  {Ferguson}, {Dickinson}, \& {Ouchi}}]{Lee:2006}
{Lee}, K., {Giavalisco}, M., {Gnedin}, O.~Y., {Somerville}, R.~S., {Ferguson},
  H.~C., {Dickinson}, M., \& {Ouchi}, M. 2006, \apj, 642, 63

\bibitem[{{Lehmer} {et~al.}(2005){Lehmer}, {Brandt}, {Alexander}, {Bauer},
  {Schneider}, {Tozzi}, {Bergeron}, {Garmire}, {Giacconi}, {Gilli}, {Hasinger},
  {Hornschemeier}, {Koekemoer}, {Mainieri}, {Miyaji}, {Nonino}, {Rosati},
  {Silverman}, {Szokoly}, \& {Vignali}}]{Lehmer:2005}
{Lehmer}, B.~D., {Brandt}, W.~N., {Alexander}, D.~M., {Bauer}, F.~E.,
  {Schneider}, D.~P., {Tozzi}, P., {Bergeron}, J., {Garmire}, G.~P.,
  {Giacconi}, R., {Gilli}, R., {Hasinger}, G., {Hornschemeier}, A.~E.,
  {Koekemoer}, A.~M., {Mainieri}, V., {Miyaji}, T., {Nonino}, M., {Rosati}, P.,
  {Silverman}, J.~D., {Szokoly}, G., \& {Vignali}, C. 2005, \apjs, 161, 21

\bibitem[{{Luo} {et~al.}(2008){Luo}, {Bauer}, {Brandt}, {Alexander}, {Lehmer},
  {Schneider}, {Brusa}, {Comastri}, {Fabian}, {Finoguenov}, {Gilli},
  {Hasinger}, {Hornschemeier}, {Koekemoer}, {Mainieri}, {Paolillo}, {Rosati},
  {Shemmer}, {Silverman}, {Smail}, {Steffen}, \& {Vignali}}]{Luo:2008}
{Luo}, B., {Bauer}, F.~E., {Brandt}, W.~N., {Alexander}, D.~M., {Lehmer},
  B.~D., {Schneider}, D.~P., {Brusa}, M., {Comastri}, A., {Fabian}, A.~C.,
  {Finoguenov}, A., {Gilli}, R., {Hasinger}, G., {Hornschemeier}, A.~E.,
  {Koekemoer}, A., {Mainieri}, V., {Paolillo}, M., {Rosati}, P., {Shemmer}, O.,
  {Silverman}, J.~D., {Smail}, I., {Steffen}, A.~T., \& {Vignali}, C. 2008,
  \apjs, 179, 19

\bibitem[{{Madau}(1995)}]{Madau:1995}
{Madau}, P. 1995, \apj, 441, 18

\bibitem[{{Madau} {et~al.}(1998){Madau}, {Pozzetti}, \&
  {Dickinson}}]{Madau:1998}
{Madau}, P., {Pozzetti}, L., \& {Dickinson}, M. 1998, \apj, 498, 106

\bibitem[{{Neufeld}(1991)}]{Neufeld:1991}
{Neufeld}, D.~A. 1991, \apjl, 370, L85

\bibitem[{{Nilsson} {et~al.}(2007){Nilsson}, {M{\o}ller}, {M{\"o}ller},
  {Fynbo}, {Micha{\l}owski}, {Watson}, {Ledoux}, {Rosati}, {Pedersen}, \&
  {Grove}}]{Nilsson:2007}
{Nilsson}, K.~K., {M{\o}ller}, P., {M{\"o}ller}, O., {Fynbo}, J.~P.~U.,
  {Micha{\l}owski}, M.~J., {Watson}, D., {Ledoux}, C., {Rosati}, P.,
  {Pedersen}, K., \& {Grove}, L.~F. 2007, \aap, 471, 71

\bibitem[{{Nilsson} {et~al.}(2009){Nilsson}, {Tapken}, {M{\o}ller},
  {Freudling}, {Fynbo}, {Meisenheimer}, {Laursen}, \&
  {{\"O}stlin}}]{Nilsson:2009}
{Nilsson}, K.~K., {Tapken}, C., {M{\o}ller}, P., {Freudling}, W., {Fynbo},
  J.~P.~U., {Meisenheimer}, K., {Laursen}, P., \& {{\"O}stlin}, G. 2009, \aap,
  498, 13

\bibitem[{{Ono} {et~al.}(2009){Ono}, {Ouchi}, {Shimasaku}, {Akiyama}, {Dunlop},
  {Farrah}, {Lee}, {McLure}, {Okamura}, \& {Yoshida}}]{Ono2009}
{Ono}, Y., {Ouchi}, M., {Shimasaku}, K., {Akiyama}, M., {Dunlop}, J., {Farrah},
  D., {Lee}, J.~C., {McLure}, R., {Okamura}, S., \& {Yoshida}, M. 2009, ArXiv
  e-prints

\bibitem[{{Orsi} {et~al.}(2008){Orsi}, {Lacey}, {Baugh}, \&
  {Infante}}]{Orsi2008}
{Orsi}, A., {Lacey}, C.~G., {Baugh}, C.~M., \& {Infante}, L. 2008, \mnras, 391,
  1589

\bibitem[{{Ouchi} {et~al.}(2003){Ouchi}, {Shimasaku}, {Furusawa}, {Miyazaki},
  {Doi}, {Hamabe}, {Hayashino}, {Kimura}, {Kodaira}, {Komiyama}, {Matsuda},
  {Miyazaki}, {Nakata}, {Okamura}, {Sekiguchi}, {Shioya}, {Tamura},
  {Taniguchi}, {Yagi}, \& {Yasuda}}]{Ouchi:2003}
{Ouchi}, M., {Shimasaku}, K., {Furusawa}, H., {Miyazaki}, M., {Doi}, M.,
  {Hamabe}, M., {Hayashino}, T., {Kimura}, M., {Kodaira}, K., {Komiyama}, Y.,
  {Matsuda}, Y., {Miyazaki}, S., {Nakata}, F., {Okamura}, S., {Sekiguchi}, M.,
  {Shioya}, Y., {Tamura}, H., {Taniguchi}, Y., {Yagi}, M., \& {Yasuda}, N.
  2003, \apj, 582, 60

\bibitem[{{Ouchi} {et~al.}(2004){Ouchi}, {Shimasaku}, {Okamura}, {Furusawa},
  {Kashikawa}, {Ota}, {Doi}, {Hamabe}, {Kimura}, {Komiyama}, {Miyazaki},
  {Miyazaki}, {Nakata}, {Sekiguchi}, {Yagi}, \& {Yasuda}}]{Ouchi:2004}
{Ouchi}, M., {Shimasaku}, K., {Okamura}, S., {Furusawa}, H., {Kashikawa}, N.,
  {Ota}, K., {Doi}, M., {Hamabe}, M., {Kimura}, M., {Komiyama}, Y., {Miyazaki},
  M., {Miyazaki}, S., {Nakata}, F., {Sekiguchi}, M., {Yagi}, M., \& {Yasuda},
  N. 2004, \apj, 611, 685
  
\bibitem[{{Ouchi} {et~al.}(2005){Ouchi}, {Hamana}, {Shimasaku}, {Yamada},
  {Akiyama}, {Kashikawa}, {Yoshida}, {Aoki}, {Iye}, {Saito}, {Sasaki},
  {Simpson}, \& {Yoshida}}]{Ouchi:2005}
{Ouchi}, M., {Hamana}, T., {Shimasaku}, K., {Yamada}, T., {Akiyama}, M.,
  {Kashikawa}, N., {Yoshida}, M., {Aoki}, K., {Iye}, M., {Saito}, T., {Sasaki},
  T., {Simpson}, C., \& {Yoshida}, M. 2005, \apjl, 635, L117

\bibitem[{{Ouchi} {et~al.}(2008){Ouchi}, {Shimasaku}, {Akiyama}, {Simpson},
  {Saito}, {Ueda}, {Furusawa}, {Sekiguchi}, {Yamada}, {Kodama}, {Kashikawa},
  {Okamura}, {Iye}, {Takata}, {Yoshida}, \& {Yoshida}}]{Ouchi:2008}
{Ouchi}, M., {Shimasaku}, K., {Akiyama}, M., {Simpson}, C., {Saito}, T.,
  {Ueda}, Y., {Furusawa}, H., {Sekiguchi}, K., {Yamada}, T., {Kodama}, T.,
  {Kashikawa}, N., {Okamura}, S., {Iye}, M., {Takata}, T., {Yoshida}, M., \&
  {Yoshida}, M. 2008, \apjs, 176, 301  

\bibitem[{{Pentericci} {et~al.}(2009){Pentericci}, {Grazian}, {Fontana},
  {Castellano}, {Giallongo}, {Salimbeni}, \& {Santini}}]{Pentericci:2009}
{Pentericci}, L., {Grazian}, A., {Fontana}, A., {Castellano}, M., {Giallongo},
  E., {Salimbeni}, S., \& {Santini}, P. 2009, \aap, 494, 553

\bibitem[{{Pickles}(1998)}]{picklesyy}
{Pickles}, A.~J. 1998, \pasp, 110, 863

\bibitem[{{Pirzkal} {et~al.}(2007){Pirzkal}, {Malhotra}, {Rhoads}, \&
  {Xu}}]{Pirzkal:2007}
{Pirzkal}, N., {Malhotra}, S., {Rhoads}, J.~E., \& {Xu}, C. 2007, \apj, 667, 49

\bibitem[{{Quadri} {et~al.}(2007){Quadri}, {van Dokkum}, {Gawiser}, {Franx},
  {Marchesini}, {Lira}, {Rudnick}, {Herrera}, {Maza}, {Kriek}, {Labb{\'e}}, \&
  {Francke}}]{Quadri:2007}
{Quadri}, R., {van Dokkum}, P., {Gawiser}, E., {Franx}, M., {Marchesini}, D.,
  {Lira}, P., {Rudnick}, G., {Herrera}, D., {Maza}, J., {Kriek}, M.,
  {Labb{\'e}}, I., \& {Francke}, H. 2007, \apj, 654, 138
  
\bibitem[{{Ranalli} {et~al.}(2003){Ranalli}, {Comastri}, \&
  {Setti}}]{Ranalli2003}
{Ranalli}, P., {Comastri}, A., \& {Setti}, G. 2003, \aap, 399, 39  

\bibitem[{{Schaerer} \& {Verhamme}(2008)}]{SV:2008}
{Schaerer}, D. \& {Verhamme}, A. 2008, \aap, 480, 369

\bibitem[{{Schlegel} {et~al.}(1998){Schlegel}, {Finkbeiner}, \&
  {Davis}}]{SFD98}
{Schlegel}, D.~J., {Finkbeiner}, D.~P., \& {Davis}, M. 1998, \apj, 500, 525

\bibitem[{{Shapley} {et~al.}(2001){Shapley}, {Steidel}, {Adelberger},
  {Dickinson}, {Giavalisco}, \& {Pettini}}]{Shapley:2001}
{Shapley}, A.~E., {Steidel}, C.~C., {Adelberger}, K.~L., {Dickinson}, M.,
  {Giavalisco}, M., \& {Pettini}, M. 2001, \apj, 562, 95

\bibitem[{{Shapley} {et~al.}(2003){Shapley}, {Steidel}, {Pettini}, \&
  {Adelberger}}]{Shapley:2003}
{Shapley}, A.~E., {Steidel}, C.~C., {Pettini}, M., \& {Adelberger}, K.~L. 2003,
  \apj, 588, 65

\bibitem[{{Sheth} \& {Tormen}(1999)}]{Shetormen:1999}
{Sheth}, R.~K. \& {Tormen}, G. 1999, \mnras, 308, 119

\bibitem[{{Simon}(2007)}]{Simon2007}
{Simon}, P. 2007, \aap, 473, 711

\bibitem[{{Somerville} {et~al.}(2004){Somerville}, {Lee}, {Ferguson},
  {Gardner}, {Moustakas}, \& {Giavalisco}}]{Somerville:2004}
{Somerville}, R.~S., {Lee}, K., {Ferguson}, H.~C., {Gardner}, J.~P.,
  {Moustakas}, L.~A., \& {Giavalisco}, M. 2004, \apjl, 600, L171
  
\bibitem[{{Steidel} \& {Hamilton}(1992)}]{StHm:1992}
{Steidel}, C.~C. \& {Hamilton}, D. 1992, \aj, 104, 941  

\bibitem[{{Steidel} {et~al.}(1999){Steidel}, {Adelberger}, {Giavalisco},
  {Dickinson}, \& {Pettini}}]{Steidel:1999}
{Steidel}, C.~C., {Adelberger}, K.~L., {Giavalisco}, M., {Dickinson}, M., \&
  {Pettini}, M. 1999, \apj, 519, 1

\bibitem[{{Steidel} {et~al.}(2003){Steidel}, {Adelberger}, {Shapley},
  {Pettini}, {Dickinson}, \& {Giavalisco}}]{Steidel:2003}
{Steidel}, C.~C., {Adelberger}, K.~L., {Shapley}, A.~E., {Pettini}, M.,
  {Dickinson}, M., \& {Giavalisco}, M. 2003, \apj, 592, 728

\bibitem[{{Steidel} {et~al.}(2004){Steidel}, {Shapley}, {Pettini},
  {Adelberger}, {Erb}, {Reddy}, \& {Hunt}}]{Steidel:2004}
{Steidel}, C.~C., {Shapley}, A.~E., {Pettini}, M., {Adelberger}, K.~L., {Erb},
  D.~K., {Reddy}, N.~A., \& {Hunt}, M.~P. 2004, \apj, 604, 534

\bibitem[{{Stiavelli} {et~al.}(2001){Stiavelli}, {Scarlata}, {Panagia}, {Treu},
  {Bertin}, \& {Bertola}}]{Stiavelli2001}
{Stiavelli}, M., {Scarlata}, C., {Panagia}, N., {Treu}, T., {Bertin}, G., \&
  {Bertola}, F. 2001, \apjl, 561, L37

\bibitem[{{Tapken} {et~al.}(2007){Tapken}, {Appenzeller}, {Noll}, {Richling},
  {Heidt}, {Meink{\"o}hn}, \& {Mehlert}}]{Tapken:2007}
{Tapken}, C., {Appenzeller}, I., {Noll}, S., {Richling}, S., {Heidt}, J.,
  {Meink{\"o}hn}, E., \& {Mehlert}, D. 2007, \aap, 467, 63

\bibitem[{{Treister} {et~al.}(2004){Treister}, {Urry}, {Chatzichristou},
  {Bauer}, {Alexander}, {Koekemoer}, {Van Duyne}, {Brandt}, {Bergeron},
  {Stern}, {Moustakas}, {Chary}, {Conselice}, {Cristiani}, \&
  {Grogin}}]{Treister2004}
{Treister}, E., {Urry}, C.~M., {Chatzichristou}, E., {Bauer}, F., {Alexander},
  D.~M., {Koekemoer}, A., {Van Duyne}, J., {Brandt}, W.~N., {Bergeron}, J.,
  {Stern}, D., {Moustakas}, L.~A., {Chary}, R.-R., {Conselice}, C.,
  {Cristiani}, S., \& {Grogin}, N. 2004, \apj, 616, 123

\bibitem[{{van Dokkum}(2001)}]{vandokkum2001}
{van Dokkum}, P.~G. 2001, \pasp, 113, 1420

\bibitem[{{Venemans} {et~al.}(2005){Venemans}, {R{\"o}ttgering}, {Miley},
  {Kurk}, {De Breuck}, {Overzier}, {van Breugel}, {Carilli}, {Ford}, {Heckman},
  {Pentericci}, \& {McCarthy}}]{Venemans:2005}
{Venemans}, B.~P., {R{\"o}ttgering}, H.~J.~A., {Miley}, G.~K., {Kurk}, J.~D.,
  {De Breuck}, C., {Overzier}, R.~A., {van Breugel}, W.~J.~M., {Carilli},
  C.~L., {Ford}, H., {Heckman}, T., {Pentericci}, L., \& {McCarthy}, P. 2005,
  \aap, 431, 793
  
\bibitem[{{Verhamme} {et~al.}(2006){Verhamme}, {Schaerer}, \&
  {Maselli}}]{V:2006}
{Verhamme}, A., {Schaerer}, D., \& {Maselli}, A. 2006, \aap, 460, 397  

\bibitem[{{Verhamme} {et~al.}(2008){Verhamme}, {Schaerer}, {Atek}, \&
  {Tapken}}]{V:2008}
{Verhamme}, A., {Schaerer}, D., {Atek}, H., \& {Tapken}, C. 2008, \aap, 491, 89

\bibitem[{{Virani} {et~al.}(2006){Virani}, {Treister}, {Urry}, \&
  {Gawiser}}]{Virani:2006}
{Virani}, S.~N., {Treister}, E., {Urry}, C.~M., \& {Gawiser}, E. 2006, \aj,
  131, 2373

\bibitem[{{Zehavi} {et~al.}(2005){Zehavi}, {Zheng}, {Weinberg}, {Frieman},
  {Berlind}, {Blanton}, {Scoccimarro}, {Sheth}, {Strauss}, {Kayo}, {Suto},
  {Fukugita}, {Nakamura}, {Bahcall}, {Brinkmann}, {Gunn}, {Hennessy},
  {Ivezi{\'c}}, {Knapp}, {Loveday}, {Meiksin}, {Schlegel}, {Schneider},
  {Szapudi}, {Tegmark}, {Vogeley}, \& {York}}]{Zehavi:2005}
{Zehavi}, I., {Zheng}, Z., {Weinberg}, D.~H., {Frieman}, J.~A., {Berlind},
  A.~A., {Blanton}, M.~R., {Scoccimarro}, R., {Sheth}, R.~K., {Strauss}, M.~A.,
  {Kayo}, I., {Suto}, Y., {Fukugita}, M., {Nakamura}, O., {Bahcall}, N.~A.,
  {Brinkmann}, J., {Gunn}, J.~E., {Hennessy}, G.~S., {Ivezi{\'c}}, {\v Z}.,
  {Knapp}, G.~R., {Loveday}, J., {Meiksin}, A., {Schlegel}, D.~J., {Schneider},
  D.~P., {Szapudi}, I., {Tegmark}, M., {Vogeley}, M.~S., \& {York}, D.~G. 2005,
  \apj, 630, 1

\end{thebibliography}
